\documentclass[times]{article}
\usepackage[a4paper, total={6in, 8in}]{geometry}
\usepackage{moreverb}
\usepackage[colorlinks,bookmarksopen,bookmarksnumbered,citecolor=red,urlcolor=red]{hyperref}
\newcommand\BibTeX{{\rmfamily B\kern-.05em \textsc{i\kern-.025em b}\kern-.08em
T\kern-.1667em\lower.7ex\hbox{E}\kern-.125emX}}

\usepackage{amssymb}
\usepackage{amsmath}
\usepackage{graphicx}
\usepackage{color}
\usepackage{enumerate}

\newenvironment{Smallmatrix}[1]
  {\arraycolsep=2pt\footnotesize
   \array{#1}}
  {\endarray}
\newcommand{\ssmat}[1]{\left[\begin{Smallmatrix}{#1}}
\newcommand{\tamss}{\end{Smallmatrix}\right]}
\newcommand{\spmat}[1]{\left(\begin{Smallmatrix}{#1}}
\newcommand{\tamps}{\end{Smallmatrix}\right)}

\newcommand{\smat}[1]{\left[\!\begin{array}{#1}}
\newcommand{\tams}{\end{array}\!\right]}
\newcommand{\pmat}[1]{\left(\!\!\!\begin{array}{#1}}
\newcommand{\tamp}{\end{array}\!\!\! \right)}
\newcommand{\semat}[1]{\begin{Smallmatrix}{#1}}
\newcommand{\tames}{\end{Smallmatrix}}
\newcommand{\emat}[1]{\!\begin{array}{#1}}
\newcommand{\tame}{\end{array}\!}
\newcommand{\mcal}[1]{\mathcal{#1}}
\newcommand{\mbb}[1]{\mathbb{#1}}
\newcommand{\mfrk}[1]{\mathfrak{#1}}

\newtheorem{remark}{Remark}

\newcommand{\hl}{\color{black}}
\newcommand{\hll}{\color{black}}

\begin{document}
\title{Model reduction for LPV systems based on approximate modal decomposition}
\author{T.~Luspay,~T.~P\'eni,~I.~G\H ozse,~Z.~Szab\'o,~B.~Vanek \\ 
~~~ \\ 
{\small Systems and Control Laboratory}  \\ 
{\small of Institute for Computer Science and Control,}  \\ 
{\small 1111 Budapest, Kende u. 13-17., Hungary} \\
{\small E-mail: \texttt{peni.tamas@sztaki.mta.hu}}}

\date{}

\maketitle

\subsection*{Abstract}
\vspace{-2pt}
The paper presents a novel model order reduction technique for large-scale linear parameter varying (LPV) systems. {\hl The approach is based on decoupling the original dynamics into smaller dimensional LPV subsystems that can be independently reduced by parameter varying reduction methods.} The decomposition starts with the construction of a modal transformation that separates the {\hl modal subsystems}. Hierarchical clustering is applied then to collect the dynamically similar {\hl modal subsystems into larger groups. The subsystems formed from the groups are then independently reduced.} This approach substantially differs from most of the previously proposed LPV model reduction techniques, {\hll since it performs manipulations on the LPV model and not on a set of  linear time-invariant (LTI) models defined at fixed scheduling parameter values. Therefore the model interpolation, which is the most challenging part of most reduction techniques, is avoided.} The applicability of the developed algorithm is thoroughly investigated and demonstrated by numerical case studies. \\

\textbf{Keywords:} linear parameter-varying systems, model order reduction, balanced realization, modal transformation, clustering

\section{Introduction}
\vspace{-2pt}

The LPV models have proven to be useful in system analysis and control design, because they are able to represent a wide class of nonlinear systems while preserving the advantageous properties of the linear time invariant (LTI) structure \cite{peni16_ijrnc}, \cite{bokor04_automatica}, \cite{pfifer15_ijrnc}. Using LPV models the analysis and control synthesis tasks are generally casted to convex optimization problems involving linear matrix inequality (LMI) constraints \cite{fenwu95_phd}. As long as the complexity of the system is low (the dimension of the state is smaller than 20-30, the number of scheduling variables is at most 2 or 3) these problems can be efficiently solved by off-the-shelf semidefinite solvers. On the other hand, the modeling of complex systems (e.g. flexible structures \cite{moreno14_ja}) results in high dimensional LPV systems (even with 100-1000 states), {\hl the dimension of which has} to be reduced in order to make them numerically tractable. {\hl It is important to emphasize that our aim is to reduce the \emph{dimension of the state vector}  and not of the scheduling variable. The latter problem is fundamentally different and is addressed e.g. in papers \cite{kwiatkowski08_ieeetcst}, \cite{rizvi16_ieeetcst}.}

 Model order reduction for LTI systems is a well-studied topic, see e.g. reference \cite{antoulas05_siam}. The same problem for LPV systems was first addressed in \cite{wood95_phd}, \cite{wood96_cdc}, where the concept of \emph{balanced realization based model truncation} \cite{moore81_ieeetac} was extended to parameter varying systems. The approach is based on the {\hl balancing state transformation computed from the}  parameter-varying controllability and observability Gramians. The computation of {\hl the Gramians} involves LMI optimization, which suffers from the same computational limitations as the LPV analysis and synthesis problems. Therefore, this model reduction  technique can only be applied to systems of moderate complexity. In order to avoid these limitations an approximate balanced truncation method is proposed in \cite{theis16_acc}. The approach is based on reformulating the balanced truncation as a Petrov-Galerkin (oblique) projection \cite{antoulas05_siam} that involves two parameter-dependent transformation matrices. These transformations depend again on the controllability and observability Gramians, but instead of computing them by LMI optimization, {\hl   \cite{theis16_acc}} proposes to directly approximate the transformation matrices by using the local Gramians associated with the LTI systems obtained at frozen scheduling parameter values.  Since the algorithm uses only QR factorization and singular value decomposition (SVD) it is numerically attractive. On the other hand, it works only for stable systems and the approximation rises some technical questions. Finding answers to these questions is part of the ongoing research.

In the literature some other model reduction methods can also be found for LPV systems, see e.g. \cite{poussotvassal12_cep}, \cite{theis16_aiaa}, \cite{adegas13_cca}. Although these techniques are different, but share a common feature; they are based on \emph{frozen parameter models}, i.e. they independently reduce the LTI models obtained at fixed scheduling parameter values and then seek a suitable interpolation algorithm to construct {\hl an LPV system from the reduced LTI model set}. The latter step is very difficult in general \cite{caigny14_ieeetcst}, because  the independently reduced (transformed and projected) local, LTI systems have to be transformed  into a consistent state-space representation.

Regarding parameter-dependent systems it is important to mention the family of \emph{parametric model reduction} methods, see e.g. \cite{benner15_acm}, \cite{geuss13_ecc}, \cite{panzer10_auto}, \cite{amsallem11_siam}. Though these approaches show similar characteristics to the LPV model reduction the problem they address is fundamentally different. {\hl The parametric model reduction starts from} a parameterized set of large-scale LTI systems. {\hl The systems are reduced and a model database is constructed from the obtained reduced order models.} If a parameter value is given, the corresponding reduced order model {\hl is} constructed from the stored model set by a suitably chosen interpolation algorithm. It is important to emphasize that the parameters in this framework are considered to be constant in time.  This is significantly different from the problem studied in this paper, since here the parameters change in time and thus the system to be reduced is time-varying.

The model reduction method proposed in this paper returns to the {\hl LPV reduction technique presented} in \cite{wood95_phd}, but instead of constructing a numerically tractable approximation for the balanced truncation, it decouples the large-scale system into smaller dimensional LPV subsystems that can be independently reduced by the {\hl original} algorithms of \cite{wood95_phd}. The decomposition starts with the construction of the parameter-varying modal transformation that separates the parameter varying modes of the system. Hierarchical clustering is applied then to group the modes into larger LPV subsystems, {\hl which can then be independently reduced. Since the model reduction is applied on LPV subsystems and not on frozen LTI models, the difficulties of system interpolation are avoided. Furthermore, the reduction of unstable systems can be naturally integrated in the proposed methodology}.

Although the (approximate) modal decomposition is part of our algorithm too, it is important to emphasize that the concept is fundamentally different from that of the \emph{modal truncation} methods elaborated for LTI systems \cite{antoulas05_siam}, \cite{varga95_mms}. Modal truncation aims at reducing the large-scale model by keeping only the dominant modal subsystems. This is not efficient in general, because the pole location does not necessarily indicate the contribution of the corresponding state to the input-output behavior of the overall system \cite{antoulas05_siam}. This is a reason why we propose to cluster the modal subsystems into larger groups and {\hl use a model reduction algorithm to reduce the larger subsystems obtained. }

The paper is organized as follows. In the following section the model reduction problem is formulated.  Section \ref{sec:modal} is devoted to the construction of the parameter-varying modal transformation. In Section \ref{sec:clustering} the hierarchical clustering and the balanced reduction methods are discussed. The numerical case studies are presented in Section \ref{sec:casestudies}. At the end of the paper the main results are summarized and the most important conclusions are drawn.

\vspace{-6pt}

\section{Problem formulation} \label{sec:problemform}
\vspace{-2pt}

A continuous-time LPV system can be given in state-space form as follows
\begin{equation}
G(\rho): \begin{array}{rcl}
\dot{x}(t)=A(\rho(t))x(t) + B(\rho(t))u(t)~\\ 
y(t)=C(\rho(t))x(t)+D(\rho(t))u(t), 
\end{array}
\label{eq:lpvsys}
\end{equation}
where $x:~ \mbb{R}_+\rightarrow \mathbb{R}^{n_x}$, $u:~ \mbb{R}_+\rightarrow \mathbb{R}^{n_u}$ and $y:~ \mbb{R}_+\rightarrow \mathbb{R}^{n_y}$, are respectively the state, input and output. The matrix functions $A: \mathbb{R}^{n_\rho} \rightarrow \mathbb{R}^{n_x \times n_x}$, $B: \mathbb{R}^{n_\rho} \rightarrow \mathbb{R}^{n_x \times n_u}$, $C: \mathbb{R}^{n_\rho} \rightarrow \mathbb{R}^{n_y \times n_x}$, $D: \mathbb{R}^{n_\rho} \rightarrow \mathbb{R}^{n_y \times n_u}$ are assumed to be continuous functions of the scheduling parameter {\hl $\rho:  \mbb{R}_+\rightarrow \mathbb{R}$.  We moreover assume  that both $\rho$ and $\dot\rho$ are bounded: $\rho(t)\in \Omega:=[\rho_{min},\rho_{max}]$ and $|\dot\rho(t)|\leq \delta$ for all $t$\footnote{For parameter-dependent variables we use the following notational convention: $a(\rho)$ denotes the variable $a$ as a function of the parameter $\rho$, while $a(\rho(t))$ or $a(\rho_k)$ denotes the value of the variable at the specific parameter value $\rho(t)$ or $\rho_k$.}.  (By this definition we restrict ourselves to LPV models having only 1 scheduling parameter. The case of vector valued $\rho$  will be  commented briefly in the conclusion.)}  

The LPV system \eqref{eq:lpvsys} is often given by a pair $(\mcal{G}, \Gamma)$, where $\mcal{G}$ is a set of LTI models obtained by evaluating \eqref{eq:lpvsys} at the {\hl finite set of scheduling parameter values} stored in $\Gamma$. Formally,    
\begin{equation}
\begin{split}
\Gamma&=\{\rho_1=\rho_{min},\rho_2,\ldots, \rho_N=\rho_{max}\},~~ \rho_1<\rho_2<\ldots<\rho_N \mbox{ and }\\
\mcal{G}&=\left\{G_k~\left |~G_k=\ssmat{c|c}A_k&B_k\\\hline C_k & D_k\tamss,~\semat{cc}A_k=A(\rho_k),&B_k=B(\rho_k),\\C_k= C(\rho_k), & D_k=D(\rho_k)\tames\right.\right\}, k=1\ldots N. 
\end{split}
\label{eq:lpvsysg}
\end{equation}
Between two consecutive grid points linear interpolation is assumed. This grid-based representation is typical in practice, because the LPV model is often generated by trimming a nonlinear system at different operating points \cite{marcos04_jgcd}. 

The aim of the LPV model reduction is to find ${G}_{red}(\rho,\dot\rho)$ of order $n_x^{red}\ll n_x$, such that the difference between the input-output behavior of the reduced and the full order models is as small as possible.  Beside time-domain simulations, several metrics can also be used to measure the similarity of the two models, see e.g. Section \ref{sec:casestudies}. These metrics give quantitative information on the applicability of the reduced order model {\hl for model-based control design}.

\section{Approximate modal decomposition for parameter-varying systems} \label{sec:modal}
For LTI systems the concept of modal decomposition is theoretically sound. {\hl Let the LTI system be given by its state-space matrices $A,B,C,D$ and let $A$ be diagonalizable.}  Then there exists a similarity state transformation $\bar T$ such that the matrix $\bar A=\bar T^{-1}A\bar T$ is block diagonal and the transformed system can be written in the form
\begin{eqnarray}
\begin{split} \label{eq:LTImodal1}
\dot{x}&=\left[
\begin{array}{cccccc}
\bar A_1 & 0 & &  \\
0 & \bar A_2 & \ddots & \\
& \ddots & \ddots & 0\\
& & 0 & \bar A_m 
\end{array}
\right]x+\left[
\begin{array}{c}
\bar B_1 \\
\bar B_2 \\
\vdots \\
\bar B_m
\end{array}
\right]u,  \\
y & =  \left[
\begin{array}{cccccc}
\bar C_1 & \bar C_2 & \ldots & \bar C_m 
\end{array}
\right]x+Du. 
\end{split}
\end{eqnarray}
where each block of the structured matrix $\bar A$ corresponds to one dynamical mode of the system, defined by the eigenvalues of $A$ as follows: {\hll  let $\lambda_1,\ldots,\lambda_{m}$ be an arbitrary sequence of the eigenvalues such that every eigenvalue is repeated as many times as its multiplicity, but the complex conjugate pairs are represented by only one member of the pair. Then} 
\begin{eqnarray}
\bar A_i = \left\{ \begin{array}{l l}
\left[
\begin{array}{cc}
\mathfrak{R}\left(\lambda_i\right) & \mathfrak{I}\left(\lambda_i\right) \\
-\mathfrak{I}\left(\lambda_i\right) & \mathfrak{R}\left(\lambda_i\right) 
\end{array}
\right] & \textnormal{if} ~~ \mfrk{I}(\lambda_i) \neq 0,\\
\left[
\lambda_i \right] & \textnormal{if} ~~ \mfrk{I}(\lambda_i) = 0
\end{array} 
\right.
\end{eqnarray}
where $\mfrk{R}(\lambda_i)$ and $\mfrk{I}(\lambda_i)$ denote the real and imaginay part of the complex eigenvalue $\lambda_i$. The similarity transformation $\bar T$ can be constructed from the {\hll eigenvectors} of $A$ as follows: {\hll let $v_1,\ldots, v_m$ be the set of distinct eigenvectors corresponding to the eigenvalue sequence above. Again, the complex conjugate eigenvector pairs are represented by only one member of the pair. Then }
\begin{equation}
\bar T = 
[\nu_1,\ldots,\nu_m],~~\mbox{where } \nu_i=\begin{cases} \mfrk{R}(v_i) & \mbox{if } \mfrk{I}(v_i)=0 \\ \left[\mfrk{R}(v_i) ~~ \mfrk{I}(v_i)\right] & \mbox{otherwise} \end{cases} 
\label{eq:T_mod}
\end{equation}
Note that the modal transformation is not unique, because it can be constructed from any eigenvector-set spanning the same eigen-space. 

Our aim is to extend this concept to parameter-varying systems and construct a parameter-dependent state transformation  $\bar{T}(\rho)$ that is differentiable in $\rho$ and decouples (at least approximately) the LPV system into modal subsystems. For this, we start from the {\hl grid-based representation \eqref{eq:lpvsysg} and compute first the} eigen-decomposition of the $A_k$ matrices. {\hl  Then an algorithm is constructed} for shaping the eigenvectors at each grid point such that the generated modal transformation sequence $\bar T_1,\ldots \bar T_N$ becomes suitable for smooth interpolation in $\rho$.

\subsection{Assumptions}
\label{subsec:assumptions}
In order to proceed, we make the following assumptions on the full order LPV model:
\begin{enumerate}[(A)]
\item \label{ass:analytic} $A(\rho)$ is diagonalizable and has a differentiable eigenvalue decomposition, i.e. there exist a diagonal $\Lambda(\rho)$ and invertible $V(\rho)$ matrices such that both are differentiable in $\rho$ and $A(\rho)=V(\rho)\Lambda(\rho) V(\rho)^{-1}$. The diagonal entries $\lambda_i(\rho)$ of $\Lambda(\rho)$ are the parameter-dependent eigenvalues, while the columns $v_i(\rho)$ of $V(\rho)$ give the parameter-dependent eigenvectors. 
\item  \label{ass:eigmult} The multiplicity of each $\lambda_i(\rho)$ is constant over the parameter domain.
\item  \label{ass:imagaxis} The $A_k$ matrices do not have eigenvalues on the imaginary axis.
\end{enumerate}
Assumption \eqref{ass:analytic} {\hl implies first that each $A_k$ matrix is} diagonalizable, so the eigen-decomposition $\Lambda_k=V_k^{-1}A_k V_k$ exists in every grid point. Second, the analyticity of $V(\rho)$ makes it possible to construct a differentiable and invertible parameter-varying state transformation from the  eigenvectors $V_k$. { \hl Matrices satisfying assumption \eqref{ass:analytic}  have been subject of intensive research for decades, see e.g. \cite{lancaster64_nm}, \cite{lancaster03_siam}, \cite{quian15_siam}. These papers provide rigorous mathematical conditions for the existence of an analytic eigenvalue/eigenvector system.} However, these conditions are rather theoretical and as explained next, they can be relaxed in the most engineering applications. Assumption \eqref{ass:eigmult} is rather technical, it guarantees that the dimension of the eigenspace associated with each $\lambda_i(\rho_k)$ is constant for all $\rho_k\in\Gamma$. This makes it easier to transform the eigenvectors into a smooth sequence, from which a differentiable state transformation can be constructed. Assumption \eqref{ass:imagaxis} is  {\hl necessary}, because the hyperbolic metric introduced in Section \ref{subsec:hyperbolic} is not defined for eigenvalues on the imaginary axis. Note that assumption \eqref{ass:imagaxis} does not exclude the case, when an eigenvector crosses the imaginary axis at some $\bar\rho$ value. If the parameter grid is chosen such that  $\bar\rho\notin\Gamma$, the proposed algorithms remain applicable.        

Though the assumptions above seem restrictive, they can be easily satisfied in the most real applications. This is because the LPV formalism is only an approximation of the underlying physical system, hence it is possible to apply minor numerical modifications in the system’s description, as long as the overall input/output behavior remains almost uneffected. Therefore, small perturbations on the $A(\rho)$ matrix is allowed as long as they do not significantly influence the {\hl input/output map realized by the dynamical system}. In real applications this is generally sufficient to satisfy the assumptions above. For further details on these possible numerical corrections see Remark \ref{rem:ncorrections} and the numerical example in Section \ref{subsec:benchmark}.

\subsection{Eigen-decomposition}
If assumption \eqref{ass:analytic} holds, then each $A_k$ in \eqref{eq:lpvsysg} is diagonalizable, hence the following eigen-decomposition can be computed for each $k$:
\begin{align}
A_k \rightarrow V_k^{-1} A_k V_k =: \Lambda_k. \label{eq:eigendec}
\end{align}
Here $\Lambda_k=\mbox{blockdiag}(\lambda_{k,1},\ldots,\lambda_{k,n_x} )$ collects the {\hl eigenvalues and $V_k=[v_{k,1},\ldots,v_{k,n_x}]$ stores the \emph{normalized} eigenvectors of $A_k$}. Note that the ordering of the eigenvalues may vary over the grid points, so the sequence $\lambda_{1,i},\ldots,\lambda_{N,i}$ may not correspond to $\lambda_i(\rho_1),\ldots, \lambda_i(\rho_N)$, where {\hl $\lambda_i(\rho)$ denotes the $i$-th parameter-dependent eigenvalue of $A(\rho)$.} In order to ensure the consistency of the $\Lambda_k$ matrices, the ordering of the eigenvalues has to be modified in each grid point.  The right ordering can be found if the eigenvalues at every grid point are correctly paired with the eigenvalues at the succeeding grid point. To find this pairing, two ingredients are needed: first, a distance metric to compare the eigenvalues and second, an algorithm to find the pairing. Moreover, the algorithm should work efficiently for large number of eigenvalues as well. {\hl In the next section a suitably distance metric is proposed, which compares the eigenvalues based on dynamic similarity.} Then the pairing problem is reformulated as finding a perfect matching in a bipartite graph. This is beneficial, because the latter problem can be solved efficiently in polynomial time.

\subsubsection*{Hyperbolic distance metric.} \label{subsec:hyperbolic} 
Our goal is to connect the series of local eigenvalues in such way that the resulting continuous trajectories correspond to the {\hl parameter-varying eigenvalues of the $A(\rho)$ matrix}. {\hl In case of an LTI system each eigenvalue represents a dynamical mode, i.e. a subsystem with particular dynamical characteristics. Extending this concept to the parameter varying case, we intend to pair two eigenvalues  if they belong to the same parameter varying subsystem.} For this purpose a metric is adopted, which measures dynamic similarity between the eigenvalues. Our approach is based on the fundamental results of system theory, which relate the location of the dominant poles to the transient response. This property is more pronounced in a discrete time setup, therefore all the stable eigenvalues are mapped into the unit circle for further discussion.  {\hl Each  $z' \in \mathbb{C}$ discrete-time eigenvalue} defines a first-order SISO transfer function $G'(z)$ with a single pole located at $z'$:
\begin{equation}
G'(z)=\frac{a'(z)}{z'-z},
\end{equation}
where $a'(z)$ is {\hl an} arbitrary polynomial in $z$. The collection of all possible transfer functions, generated by a single eigenvalue $z'$ is denoted by $\mfrk{G}[z']$. It is clear, that $\mfrk{G}[z']\subset H^2(\mbb{D})$, where $H^2(\mbb{D})$ denotes the Hardy space of complex-valued functions\footnote{which are holomorphic on the unit disk and have finite 2-norm, i.e. $\|f\|_2^2:=\frac{1}{2\pi}\int_0^{2\pi}|f(re^{i\theta})|^2d\theta<\infty$ for all $f\in H^2(\mbb{D})$} \cite{toth10_phd}. {\hl This is an infinite dimensional vector space. Among its possible orthonormal bases} the discrete Laguerre system has received distinct attention in system and control theory \cite{soumelidis11_cdcecc}. {\hl The Laguerre system is generated by any complex number $z''$ as follows}:
\begin{align} 
\label{eq:Laguerre}
\Phi''_m(z)=\frac{\sqrt{1-|z''|^2}}{1-(z'')^*\,z}\,\left(\frac{z-z''}{1-(z'')^*\,z}\right)^m  \quad (m=0,1,2\dots),
\end{align} 
Using the Laguerre system as a basis, each transfer function {\hl in $\mfrk{G}[z']$} can be represented through the linear combination of a $\Phi_m(z)$, i.e.
\begin{align} 
 G'(z)=\sum_{m=0}^\infty l_m \Phi''_m(z)
 \label{eq:Laguerre_series}
\end{align}
where $l_m$ are the corresponding linear coefficients. It follows from the results in  \cite{heuberger06_hyperbolicbook} and \cite{soumelidis13_med} that under the given conditions, {\hl the ratio of} two consecutive coefficients in \eqref{eq:Laguerre_series} is always constant and depends only on the generator {\hl elements $z''$ and the $z'$}. That is: $l_{m+1}/l_{m}=r$ for all $m=\{1,2,\ldots\}$ and $r$ is called the \emph{convergence factor}, because it characterizes the convergence of the Laguerre series \eqref{eq:Laguerre_series}.  Small coefficient implies that only a few elements are dominant in \eqref{eq:Laguerre_series}, {\hl i.e.  few basis functions are enough} to capture the dynamic behaviour of $G'(z)$. The metric $r$ is thus characterizes the similarity between the transfer functions $G'(z)$ generated by $z'$ and the basis elements parameterized by  $z''$. Therefore, it can be considered as a possible measure of the dynamic similarity between $z'$ and $z''$. Furthermore, it can be proved (see e.g. \cite{heuberger06_hyperbolicbook}) that $r$ can be computed by the following simple formula 
\begin{align} 
r=h(z',z'')=\left| \frac{z'-z''}{1-z'^*z''} \right|,
\label{eq:hypdist}
\end{align}
which is also known as the \emph{pseudo-hyperbolic metric} between $z'$ and $z''$  \cite{beardon00_iwqma}, \cite{szabo15_hyperbolicbook}.

Being defined on the unit disk, this metric can only be applied on stable, discrete eigenvalues. In order to compare continuous eigenvalues, they have to be "discretized" first by the formula $\lambda_d=\exp(\lambda_c T_s)$, where $T_s$ is a sufficiently small sampling time and $\lambda_c,\lambda_d$ denote the continuous eigenvalue and its discrete counterpart, respectively. Formula \eqref{eq:hypdist} can be applied to discrete, unstable eigenvalues as well provided that they are transformed into the unit disk 
by the mapping $f(z)=1/z^*$.  {\hll (Note that $f(z)$ reflects the unstable eigenvalue across the unit circle, which guarantees that the  distance between a stable and a (transformed) unstable eigenvalue is small if they are close to each other in the sense that they can be considered as two consecutive points of the same eigenvalue trajectory. This property is important when mixed stability eigenvalues have to be identified.)}  For notational convenience, the hyperbolic distance of two stable/unstable continuous/discrete eigenvalues will also be denoted by $h(\cdot,\cdot)$ and the transformations necessary to use \eqref{eq:hypdist} are assumed to be performed beforehand.

Note also that \eqref{eq:hypdist} is 1 if either $z'$ or $z''$ is on the unit disk.  Therefore the metric {\hl has no use} for continuous eigenvalues on the imaginary axis. {\hl This is the reason why assumption \eqref{ass:imagaxis} has been introduced in Section \ref{subsec:assumptions}}. 

Returning to our original problem, the hyperbolic distance provides a possible method to compare the eigenvalues $\lambda_{k,i}$ and $\lambda_{k+1,j}$  {\hl based on} dynamic similarity. In order to take the associated eigenvectors $v_{k,i}$ and $v_{k+1,j}$ into consideration as well, the distance metric $h(\lambda_{k,i},\lambda_{k+1,j})$ {\hl can be} weighted by $1-|v_{k,i}^*v_{k+1,j}|$, where $|v_{k,i}^*v_{k+1,j}|$ is the modal assurance criterion (MAC) \cite{ewins84_mac} that is a suitable measure for the directional similarity of the normalized eigenvectors. Denoting the weighted distance metric by $h_w(\cdot,\cdot)$ the {\hl final} formula we use to compare two eigenvalues (and their eigenvectors) can be given as follows:
\begin{align} 
h_w(\lambda_{k,i},\lambda_{k+1,j}):=h(\lambda_{k,i},\lambda_{k+1,j})\cdot (1-|v_{k,i}^*v_{k+1,j}|)
\label{eq:whypdist}
\end{align}

\subsubsection*{Perfect matching in complete bipartite graphs.} \label{subsec:hungarian} 
Let $\mcal{L}_k$ and $\mcal{L}_{k+1}$ denote the ordered sets containing the eigenvalues at grid points $k$ and $k+1$, respectively, i.e. $\mcal{L}_k=\{\lambda_{k,1},\ldots,\lambda_{k,n_x}\}$,  $\mcal{L}_{k+1}=\{\lambda_{k+1,1},\ldots,\lambda_{k+1,n_x}\}$. In order to find the pairing between these values, a graph-theoretic reformulation is given here. For this, let the weighted, complete, bipartite graph \cite{burkard09_assignmentbook} $\mcal{B}_k=(\mcal{N}_k,\mcal{E}_k,\mcal{C}_k)$ be defined with vertices $\mcal{N}_k=\mcal{L}_k \cup \mcal{L}_{k+1}$, edges $\mcal{E}_k=\{e_{ij}~|~e_{ij}:=(\lambda_{k,i},\lambda_{k+1,j}),~\forall (i,j) \mbox{ pairs}\}$ and edge-costs $\mcal{C}_k=\{c_{ij}~|~c_{ij}=h_w(\lambda_{k,i},\lambda_{k+1,j})\}$. So $\mcal{B}_k$ is defined such that its edges connect every element in $\mcal{L}_k$  with all elements in $\mcal{L}_{k+1}$ and the cost of an edge characterizes the dynamical similarity between the two eigenvalues on the edge. Note that, every $\lambda_{k,i}$ has exactly one pair in $\mcal{L}_{k+1}$ and every $\lambda_{k+1,j}$ is a pair of exactly one vertex in $\mcal{L}_{k}$. In terms of the graph-theoretic setup, this is a pairing between $\mcal{L}_k$ and $\mcal{L}_{k+1}$, which is a \emph{perfect matching} (\cite{lovasz86_matchingbook}, \cite{burkard09_assignmentbook}) in $\mcal{B}_k$, characterized by a set of $n_x$ independent edges in $\mcal{E}_k$. The cost of a perfect matching is defined by the sum of the costs for the corresponding set of independent edges. Accordingly, finding the right pairing between the eigenvalues of $\mcal{L}_k$ and $\mcal{L}_{k+1}$  is formulated as finding the \emph{minimum cost perfect matching} in  $\mcal{B}_k$ \cite{lovasz86_matchingbook}. The obtained matching problem can be efficiently solved in polynomial time by using the Hungarian Method (Kuhn-Munkres Algorithm) \cite{burkard09_assignmentbook}, offering a numerically attractive solution for the eigenvalue pairing problem. Considering the ordering of the eigenvalues at $\rho_1$ as a reference, the pairing problem can be solved successively for $k=1,2,\ldots, N-1$. As a result the $\mcal{L}_k$ sets (and consequently the $\Lambda_k$ matrices) {\hl will be} consistently ordered. {\hl This ordering has to be applied to the columns of $V_k$ as well, in order that the eigenvectors become consistent with the reordered eigenvalues.}

\subsection{Continuity of the eigenvectors and the Procrustes-problem} \label{subsec:procrustes} 
{\hl The next step is to follow the idea of modal decomposition and construct (by using \eqref{eq:T_mod}) a parameter varying state transformation from the eigenvectors $V_k$, $k=1\ldots N$. For this, a sequence of local modal transformations $\bar T_1,\ldots \bar T_N$ is generated from the eigenvectors and then these matrices are interpolated along the scheduling parameter.  
In order to facilitate the interpolation, the eigenvectors have to be shaped first such that their entries form a smooth function along the parameter grid.  Of course, the shaping has to preserve the eigenspace spanned by the eigenvectors}. 

To this end, the first step is to identify the multiplicity of the eigenvalues in order that the eigenvectors associated with a same eigenvalue can be grouped and handled together. For this, let $\tau_i$ denote the $i$-th eigenvalue trajectory, i.e. $\tau_i=(\lambda_{1,i},\ldots,\lambda_{N,i})$. By using  \eqref{eq:hypdist}, we can introduce a distance metric between two eigenvalue trajectories as follows\footnote{Note that, \eqref{eq:clustermetric} can also be defined by the weighted $h_w(\cdot,\cdot)$ metric as well.}: 
\begin{align}
H(\tau_i,\tau_j)=
\min \left( ~\max_k h(\lambda_{k,i},\lambda_{k,j}) ~,~ \max_k h(\lambda_{k,i},\lambda_{k,j}^*) ~ \right). \label{eq:clustermetric}  
\end{align}
 This metric does not distinguish the complex pairs, i.e. the distance of $\tau_i$ from $\tau_j$ and $\tau_j^*$ are the same. This is important to render the complex conjugate eigenvalue sequences together. By computing the distance between every ($\tau_i$, $\tau_j$) pair, the multiple eigenvalue trajectories can be identified. {\hll Two eigenvalue trajectories are considered to belong to a single repeated eigenvalue, if the distance between them is smaller than a given threshold. (This threshold-based decision is necessary here, because after the numerical manipulations (e.g. the eigen-decomposition) performed on the LPV model we might not expect that the eigenvalue sequences corresponding to a multiple eigenvalue of $A(\rho)$ will be perfectly equal in each grid point and give 0 distance. On the other hand, if the original system contains distinct eigenvalues that are very close to each other, then from numerical point of view, it is generally better  to handle them as a single, repeated eigenvalue.)}

 By assigning to every eigenvalue the eigenvectors spanning its eigenspace, an ordered set \- $\{(\lambda_{k,1},V_{k,1}),\ldots, (\lambda_{k,n},V_{k,n})\}$, $n\leq n_x$ can be defined at each grid point. Here 
$V_{k,i}\in\mbb{C}^{n_x\times d_i}$ collects the eigenvectors associated with $\lambda_{k,i}$.  Due to assumption \eqref{ass:eigmult} the dimension $d_i$ of the eigen-space is constant and the same at every grid point, for any $i$. This property will be exploited in the algorithm below. 

The next step is to transform the eigenvector sequence $V_{1,i},\ldots,V_{N,i}$ for all $i$. This can be done by a right multiplication of $V_{k,i}$ with an invertible matrix $Q_{k,i}$, which changes the eigenvectors, however leaves the eigenspace intact. To obtain the required smooth interpolation, $Q_{k,i}$ should transform the respective eigenvectors at consecutive grid points as close as possible. This condition is formulated as a complex, unconstrained Procrustes problem \cite{golub13_matrixbook}, \cite{geuss13_ecc}, \cite{amsallem11_siam} as follows:
\begin{align}
\bar Q_{k+1,i} := \textnormal{arg} \min_{Q_{k+1,i}} \left\|V_{k,i}-V_{k+1,i}Q_{k+1,i}\right\|_F,   ~~\forall k,i
\label{eq:procprob}
\end{align}
where $k$ goes from 1 to $N-1$, $i\in\{1,\ldots,n\}$, $Q_{k+1,i} \in \mathbb{C}^{d_i \times d_i}$, with $d_i$ denoting the dimension of the eigenspace associated with the $i$-th eigenvalue. The solution for \eqref{eq:procprob} can be analytically given in the following closed form:
\begin{align}
\mbox{vec}(\bar Q_{k+1,i})= \left(I \otimes \left[
\begin{array}{cc}
\mathfrak{R}\left(V_{k+1,i}\right) & -\mathfrak{I}\left(V_{k+1,i}\right) \\
\mathfrak{I}\left(V_{k+1,i}\right) & \mathfrak{R}\left(V_{k+1,i}\right)
\end{array}
\right]\right)^\dagger \mbox{vec}\left(\left[
\begin{array}{c}
\mathfrak{R}\left(V_{k,i}\right) \\
\mathfrak{I}\left(V_{k,i}\right)
\end{array}
\right]\right)
\end{align}
{\hll where vec($\bar Q_{k+1,i}$) is a vector formed formed from $\bar Q_{k+1,i}$ by stacking its columns below each other.} Accordingly, $\bar{V}_{k+1,i}=V_{k+1,i}\bar Q_{k+1,i}$ is the appropriately rotated eigenvector, consistent with $V_{k,i}$. The described Procrustes problem hence can be solved successively for $k=1,2,\ldots N$, for each eigenvalue of the system.  It should be noted, that the Procrustes iteration can be started from any other grid point as well.  The iteration has to be performed then in both directions: for the larger and for the smaller indices. {\hl (For choosing the starting grid point it is a reasonable approach to analyze the  numerical conditioning of the  eigenvector matrices and choose a grid point, where the associated $V_k$ matrix is well-conditioned. This improves the numerical stability and reliability of the further computations.) }

\begin{remark} \label{rem:ncorrections}
If assumption \eqref{ass:eigmult}  does not hold the Procrustes algorithm cannot be applied because the dimension of the eigenspaces changes at some  grid points. If assumption \eqref{ass:analytic} is not satisfied the Procrustes  algorithm fails to generate smooth eigenvector sequence. However, if either problem is detected it is possible in general to correct the eigenvalues and eigenvectors at the specific grid points such that the input-output behavior of the overall system does not change significantly, but after the corrections, the Procrustes algorithm can run and returns a smooth solution. To illustrate the idea, consider the case when a repeated real eigenvector trajectory switches at certain grid points to a complex pair. In real applications this is a common source of discontinuity, which is caused in general by  the numerical inaccuracies  of the eigen-decomposition. In order to resolve this issue, the complex eigenvalues can be substituted by their real part and the corresponding eigenvector pairs can be simply replaced by the eigenvectors associated with the closest real eigenvalue. This, of course, modifies the $A(\rho)$ matrix, but as long as this modification has small impact on the dynamical behavior, it is an allowed correction to eliminate the discontinuity. The case study in Section \ref{subsec:benchmark} gives an example for the application of this idea in practice.
\end{remark}

\subsection{Approximate modal transformation}
{\hl The sequence $\bar V_1,\ldots \bar V_N$  of the shaped eigenvector matrices (with $\bar V_k=[\bar V_{k,1}~\ldots~ \bar V_{k,n}]$) can be used to construct the local modal transformations $\bar T_1,\ldots,\bar T_N$}. Due to the corrections above, the entries of $\bar T_{k}$ can be smoothly interpolated (e.g. by splines) {\hl over} the parameter domain. Therefore, a parameter dependent, differentiable transformation $\bar T(\rho)$ can be obtained.  Defining a new state vector $\bar x$ such that $\bar T(\rho)\bar x = x$, the original LPV system \eqref{eq:lpvsys} transforms into  
\begin{eqnarray}
\label{eq:transdyn1} 
\begin{split}
\dot{\bar x}&= \left(\bar T^{-1}(\rho)A(\rho)\bar T(\rho)-\bar T^{-1}(\rho)\frac{\partial \bar T(\rho)}{\partial \rho}\dot{\rho}\right)\bar x
 + \bar T^{-1}(\rho)B(\rho)u \\ 
 &=(\bar A(\rho) +\bar E(\rho,\dot\rho ))\bar x+\bar B(\rho) u \\
 y&= C(\rho) \bar T(\rho) \bar x + D(\rho) u =\bar C(\rho)\bar x+ D(\rho)u.
 \end{split}
\end{eqnarray}
where {\hll $\bar A( \rho)=\bar T^{-1}(\rho)A(\rho)\bar T(\rho)$ } similarly to \eqref{eq:LTImodal1} is block diagonal with 1 or 2 dimensional blocks. $\bar E(\rho,\dot\rho)$ contains the $\dot\rho$-dependent term in \eqref{eq:transdyn1} and the difference $\bar T^{-1}(\rho)A(\rho)\bar T(\rho)-\bar A(\rho)$. The latter is zero only at the grid points.  The term $\bar E(\rho,\dot\rho)$ represents the main challenge in the application of the transformation, as noticed in the LPV literature. The grid-based representation of \eqref{eq:transdyn1} can be given as $(\bar{\mcal{G}},\Gamma \times \Omega)$, where
{\hl \begin{align} 
\bar{\mcal{G}}&=\left\{\bar G_{k,s}~\left |~\bar G_{k,s}=\ssmat{c|c}\bar A_k+\bar E_{k,s}&\bar B_k\\\hline \bar C_k &  D_k\tamss,~\semat{cc}\bar A_k=\bar A(\rho_k),&\bar B_k=\bar B(\rho_k),\\\bar C_k= \bar C(\rho_k), &  D_k=D(\rho_k)\tames\right.\right\}, \semat{cc} k=1\ldots N \\ \rho_k\in\Gamma \tames 
\label{eq:modal1}
\end{align}}
and 
{\hl \begin{align}
\bar E_{k,s}=\bar E(\rho_k,\nu_s)=-\bar T^{-1}(\rho_k)\left.\frac{\partial \bar T(\rho)}{\partial \rho}\right |_{\rho=\rho_k}\nu_s,~~ \mbox{ with } ~~ s\in\{1,2\},~ \nu_s\in\{-\delta,\delta\}
\label{eq:modal2}
\end{align}}
Note that the difference term $\bar T^{-1}(\rho)A(\rho)\bar T(\rho)-\bar A(\rho)$ is automatically neglected, because it is 0 for every $\rho_k\in\Gamma$. It is clear, that without the $\bar E(\rho,\dot\rho)$ (or $\bar E_{k,s}$) term, the transformed system is fully decoupled and is similar to the modal form \eqref{eq:LTImodal1}.  At this point the transformed model has to be analyzed in order to decide if  the $\dot\rho$-dependent term $\bar E_{k,s}$ is kept or neglected. If the original and the modal system {\hll (i.e. \eqref{eq:modal1}-\eqref{eq:modal2} without $\bar E_{k,s}$)} do not differ significantly in terms of input-output behavior, the $\bar E_{k,s}$ term can be dropped out and the computations can be proceeded with the decoupled structure.  Otherwise, it has to be kept and taken into account in the final, model reduction step discussed in Section \ref{subsec:balred}.


\section{Clustering and balanced reduction}\label{sec:clustering}
\vspace{-2pt}

The algorithm we have developed so far is able to decouple (at least approximately) the LPV system into a set of independent parameter-varying modal subsystems. In order the reduce the dimension of the overall model, the idea is to group the modal subsystems of similar dynamical properties into clusters, in order to construct larger dimensional subsystems that can be efficiently reduced.  The use of dynamic similarity  is important, because we can expect significant dimension reduction only if the subsystems are built up from similar dynamical components. 

In the forthcoming two subsections, we show first how the hierarchical agglomerative clustering (HAC) methodology \cite{manning08_clusteringbook}, \cite{hastie09_clusteringbook} can be adapted to solve our specific clustering problem. Then, based on \cite{wood95_phd}, the main steps of the parameter-varying balanced reduction algorithm are recalled.    

\subsection{Hierarchical clustering} \label{subsec:clustering}
In this section the hierarchical agglomerative clustering  methodology is adopted to group the similar {\hl modal subsystems} together.  The HAC framework is a bottom-up clustering approach, where each data object is treated as a singleton cluster and successively merged until a single cluster is obtained. 

In the context of the model reduction, the eigenvalue trajectories $\tau_i$, $i=1,\ldots,n_x$ are considered as the individual data objects. For measuring similarities between these trajectories, formula \eqref{eq:clustermetric} is used.  As this metric ensures merging of complex pairs into one cluster, the parameter-varying modes -- $(\tau_i,\tau_i^*)$ pairs -- take place at the lowest level of the HAC. For the comparison of two clusters the complete link clustering is used, i.e.: the similarity of two clusters is determined by the similarity of their most dissimilar members. Formally, if $C_m$ and $C_n$ are two clusters, then the corresponding merging criterion is:
\begin{equation}
{\hl L(C_m,C_n) = \left\{\max_{i,j} (H(\tau_i,\tau_j)), \, \tau_i \in C_m, \, \tau_j \in C_n\right\}. \label{eq:completelink}}
\end{equation}
Consequently, in the HAC framework, at each algorithmic step those two clusters are merged together for which the \eqref{eq:completelink} value is the smallest. Note that \eqref{eq:completelink} is non-local, i.e. the entire structure of the clustering can influence merge decisions. This results in a preference for compact clusters with small diameters (i.e. the most similar dynamics are grouped together) and causes sensitivity to outliers (i.e. uncommon dynamical components). The merging is repeated until all the objects have been grouped into a single cluster. This can be done in $O(n_x^2 \log n_x)$ steps \cite{manning08_clusteringbook}. {\hl The result of the HAC is visualized {\hl by} a \emph{dendrogram}, which is a tree diagram illustrating how the data objects are merged into larger clusters until the one single cluster is reached. 
{\hl The final cluster structure is obtained by cutting the dendrogram at a user-defined level of similarity. The careful choice of this threshold is important, because it determines the number and size of the clusters generated.} In the decomposition of dynamical systems the number of clusters is mainly limited by the size of the largest cluster, since the cluster size gives the dimension of the underlying dynamical system, which cannot be arbitrarily large due to the  numerical limitations of the balanced reduction algorithm to be applied in the next step. }

\begin{figure}[!]
\centering
\includegraphics[width=0.9\columnwidth]{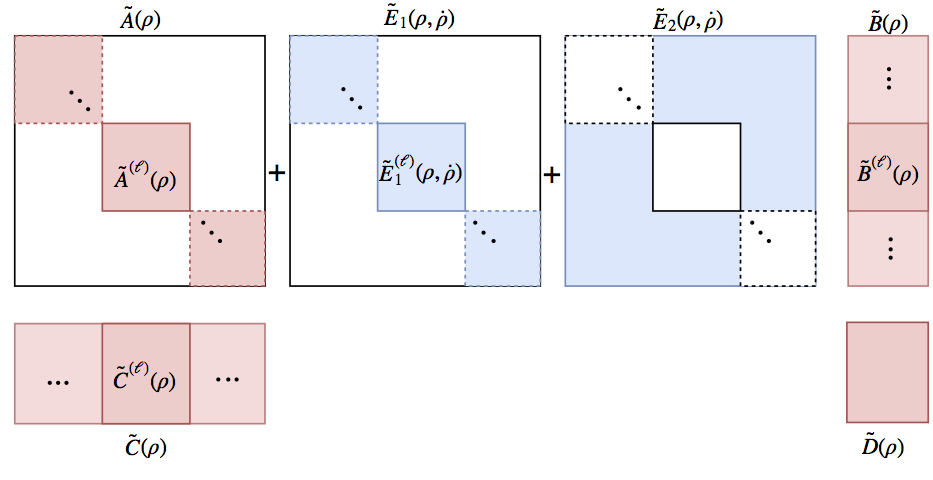}
\caption{ The structure of the system matrices after clustering: $\tilde A(\rho)=\mbox{blockdiag}(\tilde A^{(1)}(\rho),\ldots,\tilde A^{(M)}(\rho))$,  $\tilde E(\rho,\dot\rho)$ is decomposed such that $\tilde E(\rho,\dot\rho)=\tilde E_1(\rho,\dot\rho)+\tilde E_2(\rho,\dot\rho)$ such that the structure of $\tilde E_1(\rho,\dot\rho)$ is aligned with the structure of $\tilde A(\rho)$. The white areas denote the zero entries.}
\label{Fig:clusteredsys}
\end{figure}

Assume $M$ clusters have been generated. By rearranging the state $\bar x$ of \eqref{eq:transdyn1} according to these clusters the system structure presented in Fig. \ref{Fig:clusteredsys} is obtained. If the new state vector is denoted by $\tilde x$ and $\Pi$ is the permutation matrix mapping  $\tilde x$ to $\bar x$, i.e. $\bar x=\Pi \tilde x$, then 
\begin{align} 
\begin{array}{rll}
\tilde A(\rho)&=\Pi^T\bar A(\rho) \Pi, & \tilde B(\rho)=\Pi^T \bar B(\rho) \\
\tilde E(\rho,\dot\rho)&=\Pi^T \bar E(\rho,\dot\rho) \Pi & \\
\tilde C(\rho)&=\bar C(\rho)\Pi, & \tilde D(\rho)=D(\rho)
\end{array}
\end{align}
or by using the grid-based representation it can be given  by the pair $(\tilde{\mcal{G}},\Gamma\times \Omega)$, where
{\hl \begin{align} 
\tilde{\mcal{G}}=\left\{\tilde G_{k,s}~\left |~\tilde G_{k,s}=\ssmat{c|c}\tilde A_k+\tilde E_{k,s}& \tilde B_k\\\hline \tilde C_k &  D_k \tamss,~\semat{ll}\tilde A_k=\tilde A(\rho_k),&\tilde B_k=\tilde B(\rho_k),\\\tilde C_k= \tilde C(\rho_k), &  D_k= D(\rho_k)\tames\right.\right\}, ~\semat{cc} k=1\ldots N, ~s=1,2 \\ \rho_k\in\Gamma, \nu_s\in\{-\delta,\delta\}\tames 
\label{eq:lpvsysg_clustered}
\end{align}
and $\tilde E_{k,s}=\tilde E(\rho_k,\nu_s)$.}
Without $\tilde E(\rho,\dot\rho)$ the dynamics are fully decoupled into $M$ subsystems $\tilde G^{(\ell)}(\rho)=\ssmat{c|c}\tilde A^{(\ell)}(\rho) & \tilde B^{(\ell)}(\rho) \\\hline \tilde C^{(\ell)} (\rho) &  D^{(\ell)}(\rho)\tamss$, $\ell=1,\ldots,M$.  
The dimension of each subsystem equals to the size of the corresponding cluster.  The cluster size is limited by the numerical complexity of the balanced reduction algorithm, which will be performed on each subsystems next. Taking this numerical limitation into account, the maximum cluster size should not be larger than 30-40. For more details, see the discussion in the next subsection.   If we have decided earlier to neglect the coupling term $\tilde E(\rho,\dot\rho)$, then these subsystems can be handled separately. Otherwise, the coupling term has to be taken into consideration. Since a decoupled structure is needed to continue our algorithm, this is only partially possible. For this, let $\tilde E(\rho,\dot\rho)$ be expressed as a sum of $\tilde E_1(\rho,\dot\rho)$ and $\tilde E_2(\rho,\dot\rho)$  such that the structure of $\tilde E_1(\rho,\dot\rho)$ is aligned with the structure of $\tilde A$ (see Fig.  \ref{Fig:clusteredsys}). Assuming that the effect of $\tilde E_2(\rho,\dot\rho)$ on the input-output behavior is negligible, we can proceed the computations with the subsystems defined by {\hl the system matrices $\tilde A^{(\ell)}(\rho)+\tilde E_1^{(\ell)}(\rho,\dot\rho)$,  $\tilde B^{(\ell)}(\rho)$,  $\tilde C^{(\ell)} (\rho)$ and  $D^{(\ell)}(\rho)$}. These systems are also independent, so they can be separately reduced. The neglected term can be treated as a modeling uncertainty during the analysis or control synthesis procedures performed on the reduced order model. (To further minimize the approximation error {\hl it is a reasonable idea to complete the distance metric \eqref{eq:clustermetric} with an additional term penalizing large entries in  $\tilde E_2(\rho,\dot\rho)$}.)   

\subsection{Balanced reduction}\label{subsec:balred}

Balanced reduction is a fundamental approach for the model reduction of linear (time invariant and varying, as well as parameter-dependent) systems \cite{wood96_cdc}, \cite{wood95_phd}. The key concept is the balanced realization which reveals the controllability and observability properties of the system. The similarity transformation $\hat T(\rho)$, which transforms a general LPV system \eqref{eq:lpvsys} into balanced form is obtained from the observability $X_o(\rho)$ and controllability $X_c(\rho)$ Gramians\footnote{ The presented balanced reduction algorithm can only be applied to quadratically stable LPV systems. The extension of the method to unstable systems is  well documented in \cite{wood95_phd} and \cite{wood96_cdc}.}.  If the LPV system is given in a state-space form {\hl and the structure of the Gramians is a priori fixed (e.g. in the form $X_o(\rho)=X_{o,0}+\sum_{i=1}^{n_b} X_{o,i} f_i(\rho) X_{o,i}$ and $X_c(\rho)=X_{c,0}+\sum_{i=1}^{n_b} X_{c,i} g_i(\rho) X_{c,i}$ where $f_i(\rho)$, $g_i(\rho)$ are fixed basis functions and $X_{o,i}$, $X_{c,i}$, $i=1,\ldots n_b$ are free variables), then $X_o(\rho)$ and $X_c(\rho)$ can be obtained as a result of the following optimization problem \cite{wood95_phd}:
\begin{equation}
\begin{split} 
\label{eq:wooditeration}
\min_{X_{o,i},X_{c,i},i=1\ldots n_b} \sum_k \mbox{trace}~{X_o(\rho_k)X_c(\rho_k)}\quad\quad\quad\quad\quad\quad\quad\quad\quad\quad \quad\\
\left.\frac {dX_o(\rho)}{d\rho}\right|_{\rho=\rho_k}\nu_s + A(\rho_k)^TX_o(\rho_k)+X_o(\rho_k)A(\rho_k)+C(\rho_k)^TC(\rho_k) \prec 0  \\
-\left.\frac{dX_c(\rho)}{d\rho}\right|_{\rho=\rho_k} \nu_s+ A(\rho_k)X_c(\rho_k)+X_c(\rho_k)A(\rho_k)^T+B(\rho_k)B(\rho_k)^T \prec 0  \\
X_o(\rho_k)\succ 0, X_c(\rho_k)\succ 0, \forall k \in \{1,\ldots,N\}, ~\rho_k\in\Gamma \mbox{ and } s\in\{1,2\},\nu_s\in\{-\delta,\delta\}
\end{split}
\end{equation}
This is a nonconvex optimization problem, but if either $X_o(\rho)$ or $X_c(\rho)$ is fixed,  then the cost function becomes linear in the remaining variables, hence the problem reduces to a linear optimization problem with Linear Matrix Inequality (LMI) constraints. As suggested in the literature by alternately fixing $X_o(\rho)$ and $X_c(\rho)$ a numerically tractable iterative algorithm is obtained, where an initial $X_o(\rho)$ (or $X_c(\rho)$) can be calculated from the point-wise (at fixed $\rho_k$ values) time-invariant solutions. }

Although the above modifications ease the computational burden, the approach still suffers from the curse of dimensionality. The number of decision variables involved, and hence the numerical complexity grows with the state-dimension of the system.  If the number of states goes over 30-40, the LMI optimization problem becomes intractable by the off-the-shelf semidefinite solvers. The decomposition of the system into smaller, independent subsystems,  offers a remedy for this problem. 

Starting from the decoupled structure \eqref{eq:lpvsysg_clustered} the next step is to reduce the $M$ subsystems separately by the parameter varying balanced reduction method.   After computing the controllability and observability Gramians $X_o^{(\ell)}(\rho)$ and $X_c^{(\ell)}(\rho)$ for every subsystem, $\ell=1,\ldots M$, the parameter-varying balanced transformations are constructed in the following way. First the unique Cholesky factors $R_o^{(\ell)}(\rho)$ and $R_c^{(\ell)}(\rho)$ {\hll (both are analytic in $\rho$)} are determined by the algorithm given in the proof of Theorem 7.8.1 in \cite{wood95_phd}:
\begin{equation} 
\begin{split}
X^{(\ell)}_o(\rho)=\left(R_o^{(\ell)}(\rho)  \right)^T \cdot R_o^{(\ell)}(\rho),~~~\mbox{ such that }R_o^{(\ell)}(\rho)~\mbox{is upper triangular}\\
X^{(\ell)}_c(\rho)=R_c(\rho)\left(R_c^{(\ell)}(\rho)\right)^T,~~~\mbox{ such that }R_c^{(\ell)}(\rho)~\mbox{is lower triangular}
\end{split}
\end{equation}
Then a parameter-varying singular value decomposition is performed on the product $R_o^{(\ell)}(\rho)R_c^{(\ell)}(\rho)$:
\begin{align} 
\label{eq:svd}
U^{(\ell)}(\rho)S^{(\ell)}(\rho)\left(V^{(\ell)}(\rho)\right)^T=R^{(\ell)}_o(\rho)R^{(\ell)}_c(\rho)
\end{align}
where $S^{(\ell)}(\rho)$ is the diagonal matrix of the parameter-dependent singular values $\sigma^{(\ell)}_j(\rho)$ and $S^{(\ell)}(\rho)$, $U^{(\ell)}(\rho)$ and $V^{(\ell)}(\rho)$ are all analytic in $\rho$. The analyticity is ensured by computing  \eqref{eq:svd} by the Analytic Singular Value Decomposition  (ASVD) algorithm proposed in \cite{mehrmann93_etna}. 
The balancing transformation $\hat{T}^{(\ell)}(\rho)$ for the $\ell$-th subsystem can be computed as 
\begin{align} 
 \hat{T}^{(\ell)}(\rho)=R_c^{(\ell)}(\rho)V^{(\ell)}(\rho)\left(S^{(\ell)}(\rho)\right)^{-\frac{1}{2}}.
\end{align}
The balancing transformations can be applied on the subsystems and the states corresponding to small singular values can be  eliminated. Since $\hat{T}^{(\ell)}(\rho)$ transformations are parameter-dependent, the reduced systems explicitly depend on $\dot\rho$ as well \cite{wood95_phd}. Having reduced the subsystems individually, the reduced dynamics are finally joined together to obtain the low dimensional approximation of \eqref{eq:lpvsysg}.

\section{Case studies} \label{sec:casestudies}
\vspace{-2pt}

\subsection{A benchmark example}\label{subsec:benchmark}

\begin{figure}
\centering
\includegraphics[width=0.95\columnwidth]{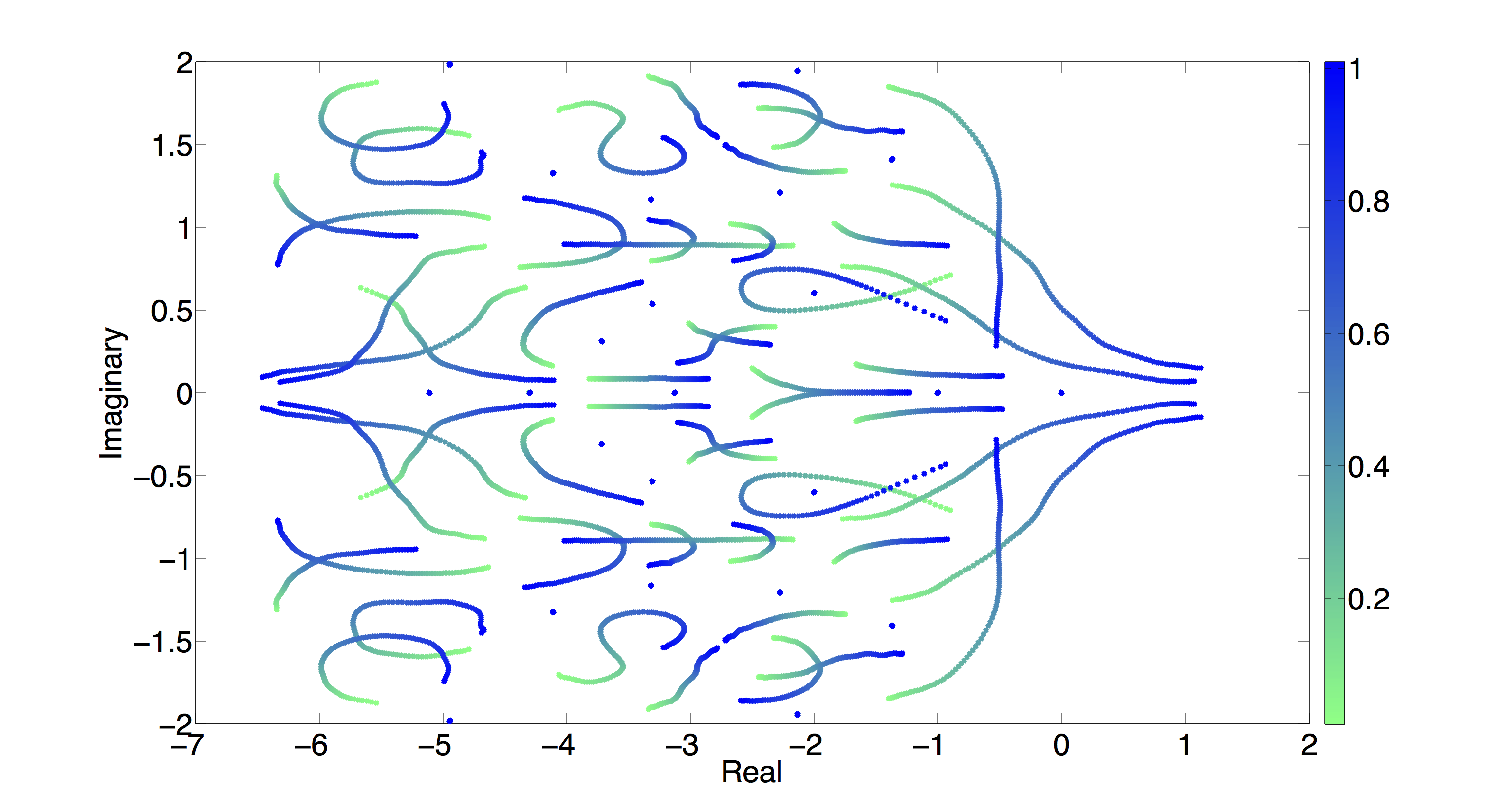}
\caption{Pole migration of the LPV benchmark system. The value of the scheduling parameter is indicated by colors according to the color-bar on the right. \label{Fig:polesys}}
\end{figure}

{\hl As a first example a $80$ dimensional, 2-input-2-output  LPV system has been generated for the numerical evaluation of the developed algorithm. The procedure used to generate the model consists of 4 steps. First, the parameter-variation of each eigenvalue is defined over the parameter domain $\Omega=[0,1]$.  {\hl The $\lambda_i(\rho)$ functions obtained are then used to construct the parameter-varying block-diagonal $A_0(\rho) \in \mathbb{R}^{80 \times 80}$ matrix. This is then completed with a randomly generated} constant input $B_0 \in \mathbb{R}^{80 \times 2}$ and output $C_0 \in \mathbb{R}^{2 \times 80}$ mappings, as well as a constant direct feedthrough term $D_0 \in \mathbb{R}^{2 \times 2}$.   To make the problem more realistic and industrially relevant, the generated modal system  is transformed in the second step by a parameter-varying matrix $T(\rho)$. $T(\rho)$ is constructed by randomly generating an invertible matrix and making its certain (randomly selected) blocks parameter-dependent. The resulting LPV system can be given as follows:
\begin{eqnarray}
G_0(\rho) : \left\{
\begin{array}{l}
\dot{x}=T(\rho)^{-1}A_0(\rho)T(\rho)x+T(\rho)^{-1}B_0u \\
y = C_0T(\rho)x + D_0 u
\end{array}
\right.
\end{eqnarray} 
The third step is evaluating the system above over an equidistant grid $\Gamma_0$ containing $N_0$ parameter values. Then a $d$-th degree polynomial is fitted to each entry of the system matrices in order to deform the eigenvalue trajectories and thus make the model more realistic. Finally, in the fourth step, a grid-based representation is generated by evaluating the model over the grid $\Gamma$. This way, the LPV system is free from any special structure and can be considered as a generic parameter-varying model.  To obtain the particular model used in this section the length $N_0$ of the initial grid, the length $N$ of the final grid and the degree $d$ of the interpolating polynomials have been chosen to be $60$, $100$ and $14$, respectively.  Furthermore, to evaluate and challenge the algorithm, a wide range of dynamical behaviors have been covered including:}
{\hl \begin{itemize}
\item parameter varying real- and complex conjugate eigenvalues,
\item parameter independent dynamics,
\item higher order complex and real eigenvalues (with algebraic multiplicity),
\item integrators,
\item mixed stability eigenvalues (that cross the imaginary axis),
\item complex conjugate - real transitions.
\end{itemize}  }
These modes can be recognized on the eigenvalue trajectories depicted in Figure \ref{Fig:polesys}. 
The numerical testing of the algorithm can be performed according to the following steps:
\begin{itemize}
\item \emph{Eigen-decomposition.} Using standard  numerical methods, the eigenvalues and the corresponding eigenvectors  of the {\hl $A_k$ matrices are} computed at each grid point. 
\item \emph{Identification of integrators.} An initial ordering is carried out, based on the absolute value of the eigenvalues in order to locate and label integrators for further computations. This is a rather technical step, necessary to satisfy assumption \eqref{ass:imagaxis}. The eigenvalues corresponding to the integrators are simply skipped from the reduction procedure, and {\hl will be}  added back only at the end, when the reduced order LPV model is assembled.
\item \emph{\hl Finding the eigenvalue trajectories.} The  samples of the pointwise modes over the parameter domain are connected by the Hungarian Algorithm, based on their weighted hyperbolic distance. Unstable poles are projected into the unit circle during the construction of the distance matrix, as discussed in Section \ref{subsec:hyperbolic}. Continuous eigenvalue trajectories are restored as a function of the scheduling parameter. Integrators are excluded from the pairing and matched individually. 
\item \emph{Detection of multiple eigenvalue trajectories}. Eigenvalue trajectories which are in close proximity for every value of the scheduling parameter are grouped together and handled as muliple ones. In a same manner, using the hyperbolic metric between trajectories, irregular behaviours (i.e. complex-real transitions) can be detected and labelled. For example, in this particular example  there are two eigenvalues very close to each other, accompanied by occasional complex-real transitions. This represents discontinuity in the eigenspace (assumptions \eqref{ass:analytic} and \eqref{ass:eigmult} are violated), hence a numerical correction (see Remark \ref{rem:ncorrections}) is needed. The absolute variance for the corresponding poles over the scheduling parameter domain was found to be in the range of $10^{-7}$ clearly indicating a constant real pole with multiplicity of $2$.  Therefore we can replace the values with their average, and the complex eigenvectors can be replaced by the closest real eigenvectors. These steps succesfully resolve the mathematical discontinuity {\hl without sensible change in the input-output behaviour}. 
\item \emph{\hl Eigenvector smoothing.} The complex Procrustes problem is solved for the matched eigenvectors, where the subspaces of the previously labelled poles are treated accordingly.  The smoothness of the modal transformations are compared numerically, by comparing the $\dot{\rho}$ dependent terms. For this purpose, a cubic spline interpolation is applied for the eigenvectors obtained from the eigen-decomposition, and for the one obtained through the Procrustes problem. These continuous functions are then evaluated between grid points. The results are summarized in Table \ref{tab:dTdrho}, where the large entries indicate discontinuity in $\partial \bar T(\rho)/\partial \rho$, which implies that the state transformation $\bar T(\rho)$ is probably not differentiable. The results illustrate the effectiveness of the proposed complex Procrustes smoothing.
\begin{table}[!h]
\centering
\begin{tabular}{l|c|c|c|c}
& $\max_\rho \max \left\{\cdot \right\}$ & $\textnormal{mean}_\rho \max \left\{\cdot \right\}$ & $\max_{\rho} \left\| \cdot \right\|_2$ & $\textnormal{mean}_\rho \left\| \cdot \right\|_2$\\
\hline
Before Procrustes $\frac{\partial \overline{T}(\rho)}{\partial \rho}$ & 100.52 & 73.6 & 312.28 & 246.28 \\
\hline 
After Procrustes $\frac{\partial \overline{T}(\rho)}{\partial \rho}$  & 4.57 & 1.07 & 12.6 & 2.85
\end{tabular}
\caption{ Numerical comparison of the maximal elements and the matrix norms of modal transformations before- and after smoothing, over the parameter domain. The numerical values were obtained by applying the functions labelling the columns to the matrices labelling  the rows, i.e. the 1,1 entry is obtained by computing the maximal entry of $\partial \bar T(\rho)/\partial \rho$ as a function of the parameter and maximizing it over $\rho$.}
\label{tab:dTdrho}
\end{table}
\item \emph{Modal form.} Applying transformation $\bar{T}(\rho)$, the parameter-varying modal form is obtained. At this point a numerical test has to be carried out to investigate the effect of the {\hl $\dot{\rho}$-dependent} term. {\hl It can be checked} via time domain simulations that {\hl in this particular example} this term can be neglected without significant change in the input/output response. (see also Table \ref{tab:dTdrho}).  
\item \emph{Stable-unstable decomposition.} By using the continuous eigenvalue trajectories and the parameter-varying modal form, unstable (as well as mixed stability) dynamics can be separated and removed from the system. In the underlying example, five states can be separated (see Figure \ref{Fig:polesys}) and thus the model reduction is performed for the remaining 75 (stable) states. In this example, this is a reasonable step, as the number of unstable/mixed stability eigenvalue trajectories is small. If this was not the case, these eigenvalues had to be preserved, which would result in unstable subsystems after clustering. In that case the variant of the balanced reduction algorithm extended to unstable systems \cite{wood95_phd} would have to be applied. 
\item \emph{Clustering.} Hiearchical clustering of the matched eigenvalue trajectories are performed next, aiming to reveal dynamical redundancies of the system. Figure \ref{Fig:dendrogram} shows the corresponding dendrogram plot, which is used for representing the arrangement of the clusters. In Figure \ref{Fig:dendrogram} the height of each line equals to the distance between the two data objects (either eigenvalue trajectory or cluster of trajectories,  computed by \eqref{eq:completelink} ) below. {\hl Based on the dendrogram, different number of clusters can be generated and it is the task of the user to decide. To make this decision the following trade-off has to be taken into consideration}: large number of clusters corresponds to smaller sized groups, in which state elimination is generally more conservative (consider the extreme case, where each cluster contains a single mode). On the other hand, small number of clusters imply higher dimensional groups with increasing numerical burden (on the other extreme: the entire system considered as a single cluster). Taking this into account, the given system are subdivided into five clusters (see Figure \ref{Fig:dendrogram}), with the following dimensions: 26 (red), 34 (purple), 7 (blue), 6 (green) and 2 (black). Note that the last cluster cannot be reduced, since it contains a single complex-conjugate mode. According to Figure \ref{Fig:dendrogram}, a dynamic dissimilarity can be observed, implying the preservation of this mode in question.
\begin{figure}[!h]
\centering
\includegraphics[width=0.95\columnwidth]{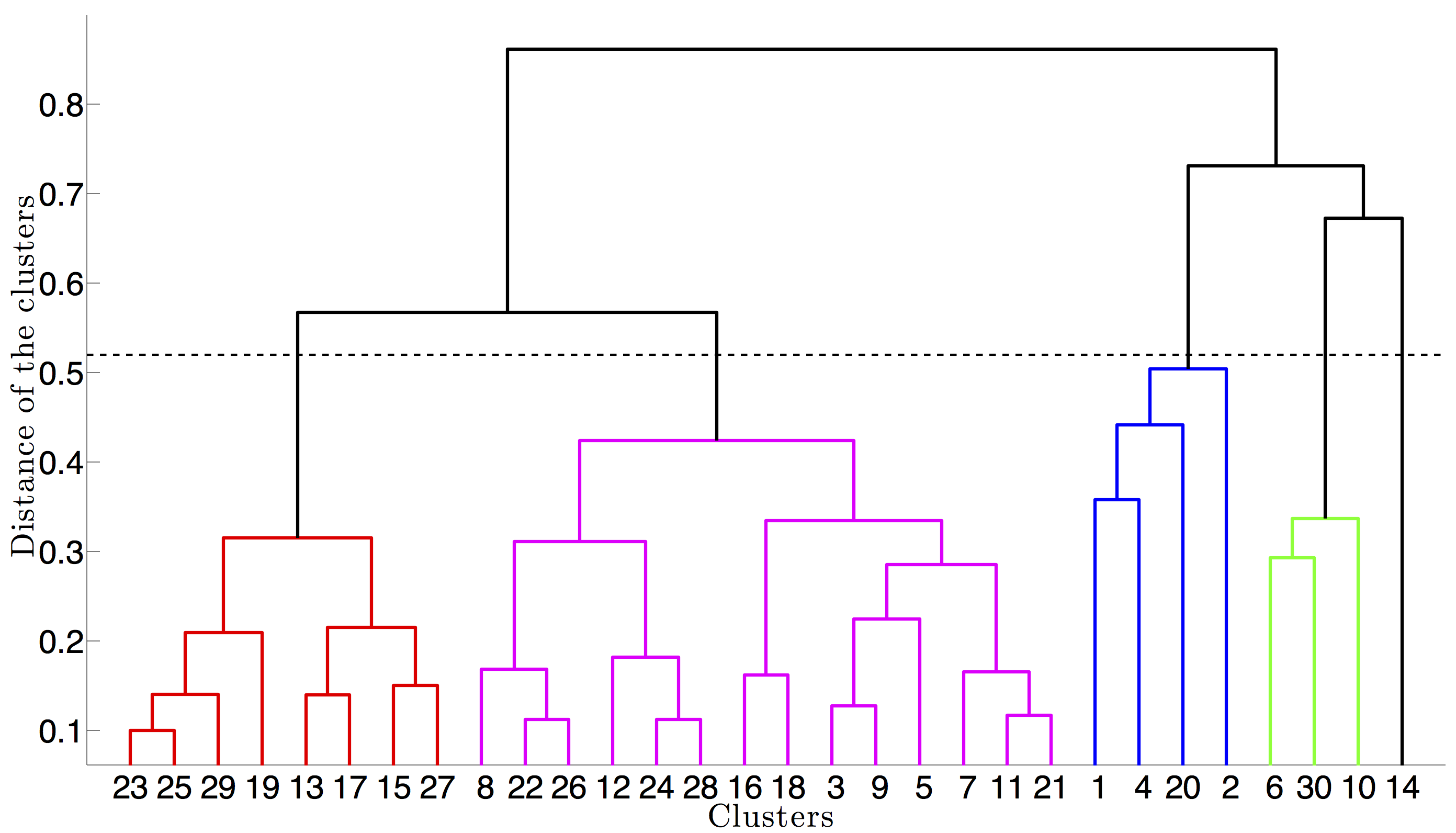}
\caption{Dendrogram for clustering in the benchmark example. The horizontal axis is labelled by the indices of the data objects (eigenvalue trajectories) while the vertical axis shows the similarity distance computed by using \eqref{eq:completelink}.  The dashed line indicates the threshold where the dendrogram was cut to obtain the 5 clusters.}
\label{Fig:dendrogram}
\end{figure}

In addition, the cophenetic correlation coefficient is computed, which is often used for characterizing a dendrogram: how faithfully does it represent the similarities among data objects  \cite{manning08_clusteringbook}. The magnitude of the cophenetic correlation should be close to $1$ for the case of a good description. The computed value was $0.83$, which shows that the selected hyperbolic metric is indeed a good indicator for comparing different dynamical behaviours. 
\item \emph{Computing Gramians.} The parameter-dependent controllability and observability Gramians are computed for each individual cluster. {\hl By fixing the structures of the Gramians as $X_c(\rho)=X_{c,0}+\rho X_{c,1}$ and $X_o(\rho)=X_{o,0}+\rho X_{o,1}$ the iterative optimization (Section \ref{subsec:balred}) is carried out for each sub-system by using the MOSEK optimization tool \cite{mosek}.} This corresponds to $(351 + 595 + 28 + 21) \times 4$ decision variables in four separate optimization problems, which is a significant decrease compared to the $5550 \times 4$ variables involved in the single LMI problem for the entire $75$ dimensional system. 
\item \emph{Analytic Singular Value Decomposition.} In order to determine the number of states with the largest contributions to the input-output behaviour of each subsystem, the ASVDs are computed individually according to \eqref{eq:svd}. {\hl The number of most significant singular values followed by the dimension of the corresponding cluster are $5/26$, $5/34$, $3/7$ and $2/6$}. 
\item \emph{Balanced Transformation.} The parameter-varying balancing transformations and their inverses can be computed for each subsystem, using the singular value decomposition \eqref{eq:svd}. Applying these transformations, balanced forms are obtained, resulting in a $\rho$, $\dot{\rho}$ dependent system.  The balanced models are then truncated according to the most significant singular values, computed in the previous step.  That is, the $75$ dimensional stable part is reduced to $17$ dimension.
\item \emph{Reduced model construction.} Finally, the individually reduced subsystems are joined together and the five dimensional unstable dynamics are added back. Hence the final, reduced-order LPV model has $22$ states (compared to the original $80$) and given in a grid-based fashion, depending on $\rho$ and $\dot{\rho}$.
\item \emph{ Evaluation of the reduced model.} In order to validate the reduced model, various numerical properties have been investigated.  One of the main motivation of model reduction is to obtain a smaller dimensional representation, which is suitable for controller design. From this point of view, similartity of closed-loop behaviours between the high-order and low-order models is of capital importance. In order to take the feedback control objective into account, the widely-used $\nu$-gap metric is evaluated, defined between LTI systems, $G_1(j\omega)$ and $G_2(j\omega)$, as \cite{georgiou88_scl}, \cite{vinnicombe93_phd}:
\begin{eqnarray}
\delta_\nu(G_1(j\omega),G_2(j\omega)) = \qquad\qquad\qquad\qquad\qquad\qquad\qquad\qquad\qquad\qquad\qquad\qquad \nonumber\\
 \left\|(I+G_2(j\omega)G_2^*(j\omega))^{-\frac{1}{2}}(G_1(j\omega)-G_2(j\omega))(I+G_1^*(j\omega)G_1(j\omega))^{-\frac{1}{2}} \right\|_\infty. \label{eq:gapdef}
\end{eqnarray} 
Essentially, if $\delta_\nu (G_1,G_2) \leq \beta$, then a controller, which stabilizes $G_2$ also stabilizes $G_1$, with a stability margin of $\beta$ \cite{vinnicombe93_phd}. For identical systems $\delta_\nu(G_1,G_2)=0$, otherwise it is $0 \leq \delta_\nu \leq 1$. For LPV systems \eqref{eq:gapdef} can be interpreted in different ways. One possibility is to compare point-wise LTI systems of the LPV dynamics by taking:
\begin{align}
\max_{\omega} \delta_\nu(G_{k}(j\omega),\tilde{G}_{k}(j\omega))
\end{align}
at each grid point $k \in [1 \, 100]$. Figure \ref{Fig:nugap} shows the evaluation of the above expression for models interpolated between the original grid points. The maximal value {\hl is} $0.12$, implying a satisfactory similarity between the full and reduced order models. 
\begin{figure}
\centering
\includegraphics[width=0.9\columnwidth]{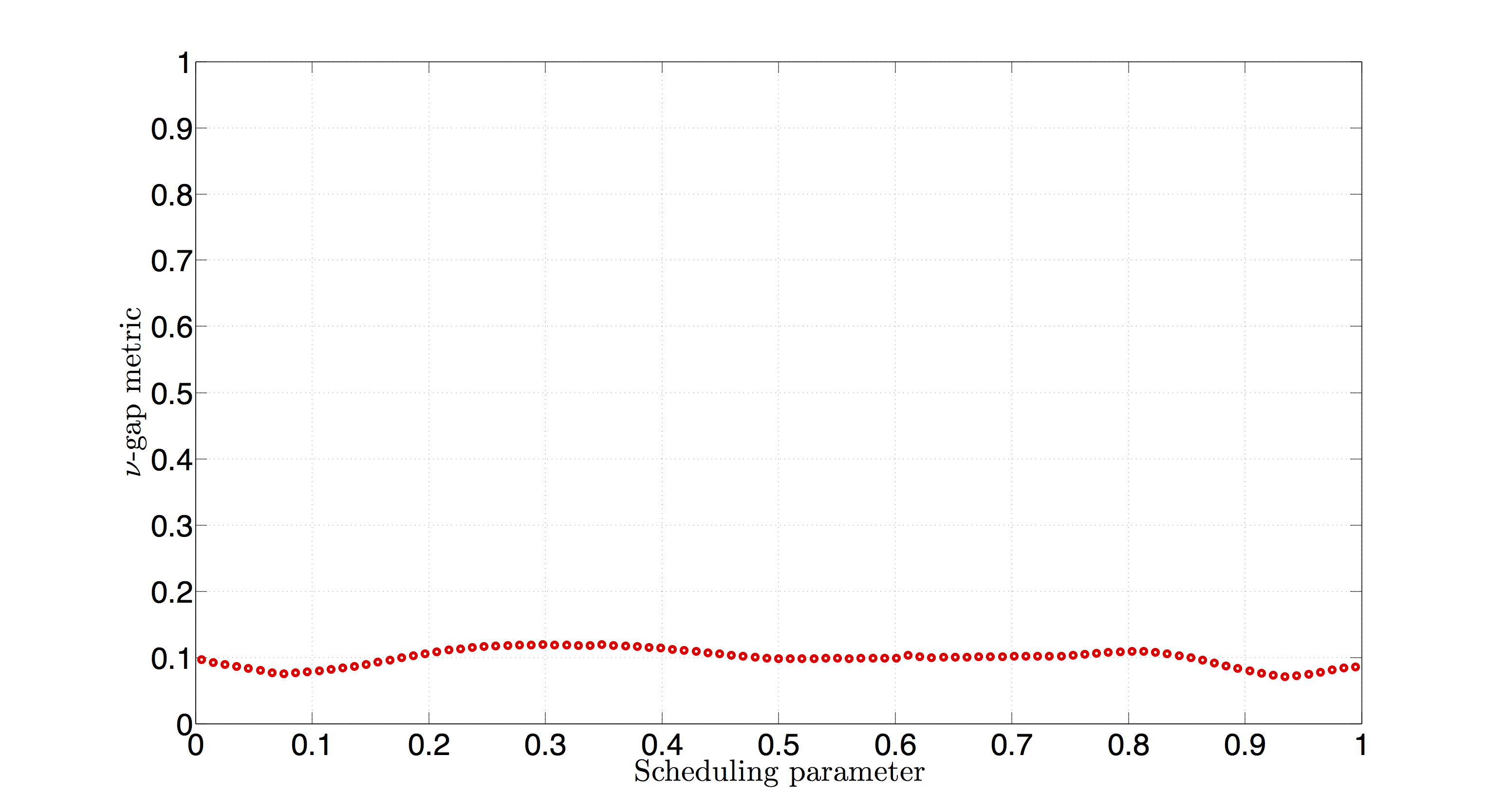}
\caption{$\nu$-gap metric of the full (80) and reduced (22) order model over the scheduling parameter domain \label{Fig:nugap}}
\end{figure}
The second representation of \eqref{eq:gapdef} is illustrated on Figure \ref{Fig:nugap_freq}, where the interpolated systems are compared at every $\omega_i$ frequency, i.e.:
\begin{align}
\max_{\rho} \delta_\nu(G_(\rho)(j\omega_i),\tilde{G}(\rho)(j\omega_i)).
\end{align}
Figure \ref{Fig:nugap_freq} also provides an insight on the frequency-domain properties of the LPV model reduction. 
\begin{figure}
\centering
\includegraphics[width=0.9\columnwidth]{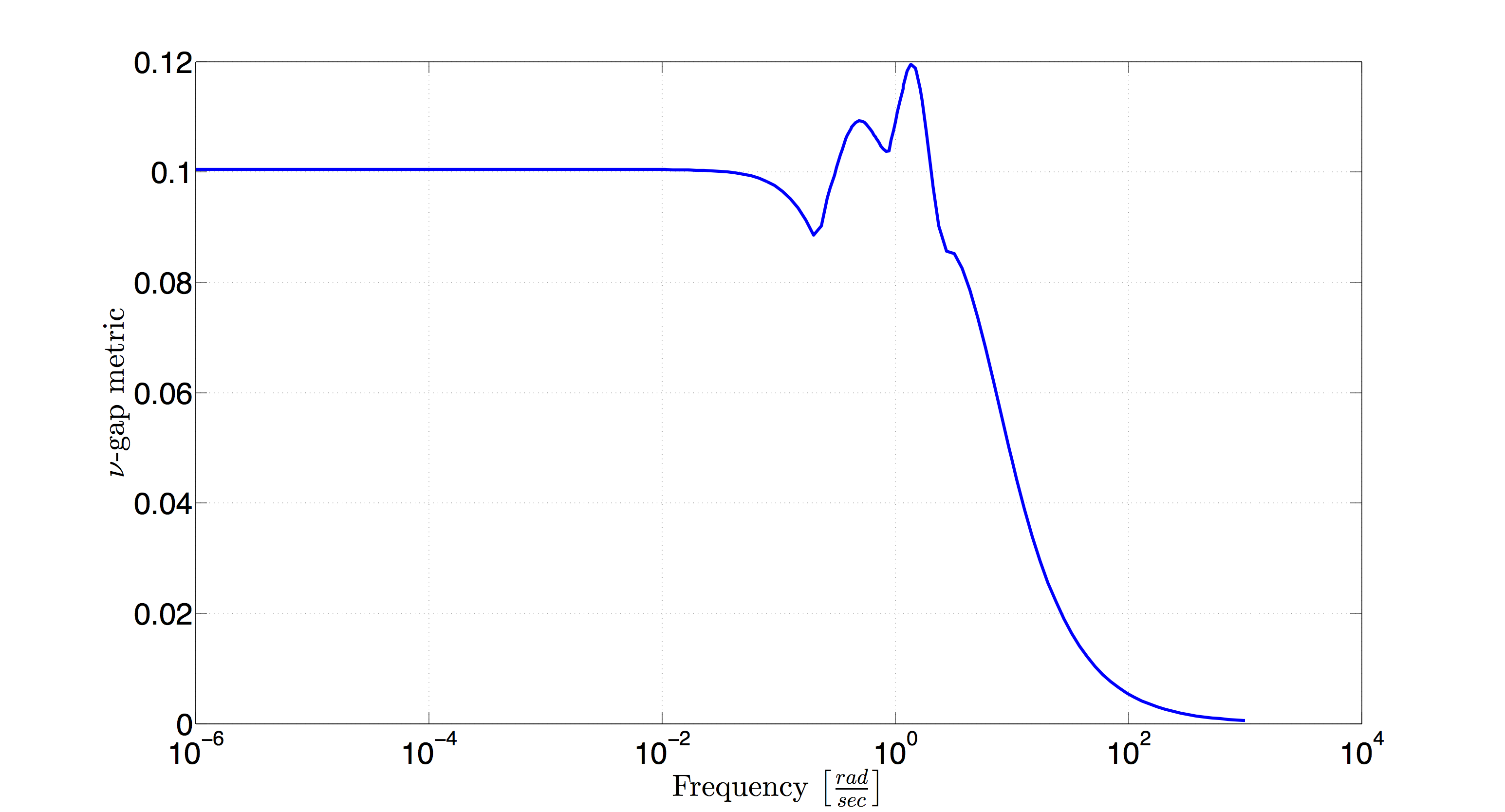}
\caption{Frequency-wise maximum $\nu$-gap metric of the full (80) and reduced (22) order model over the scheduling parameter domain \label{Fig:nugap_freq}}
\end{figure}
\begin{figure}
\centering
\includegraphics[width=\textwidth]{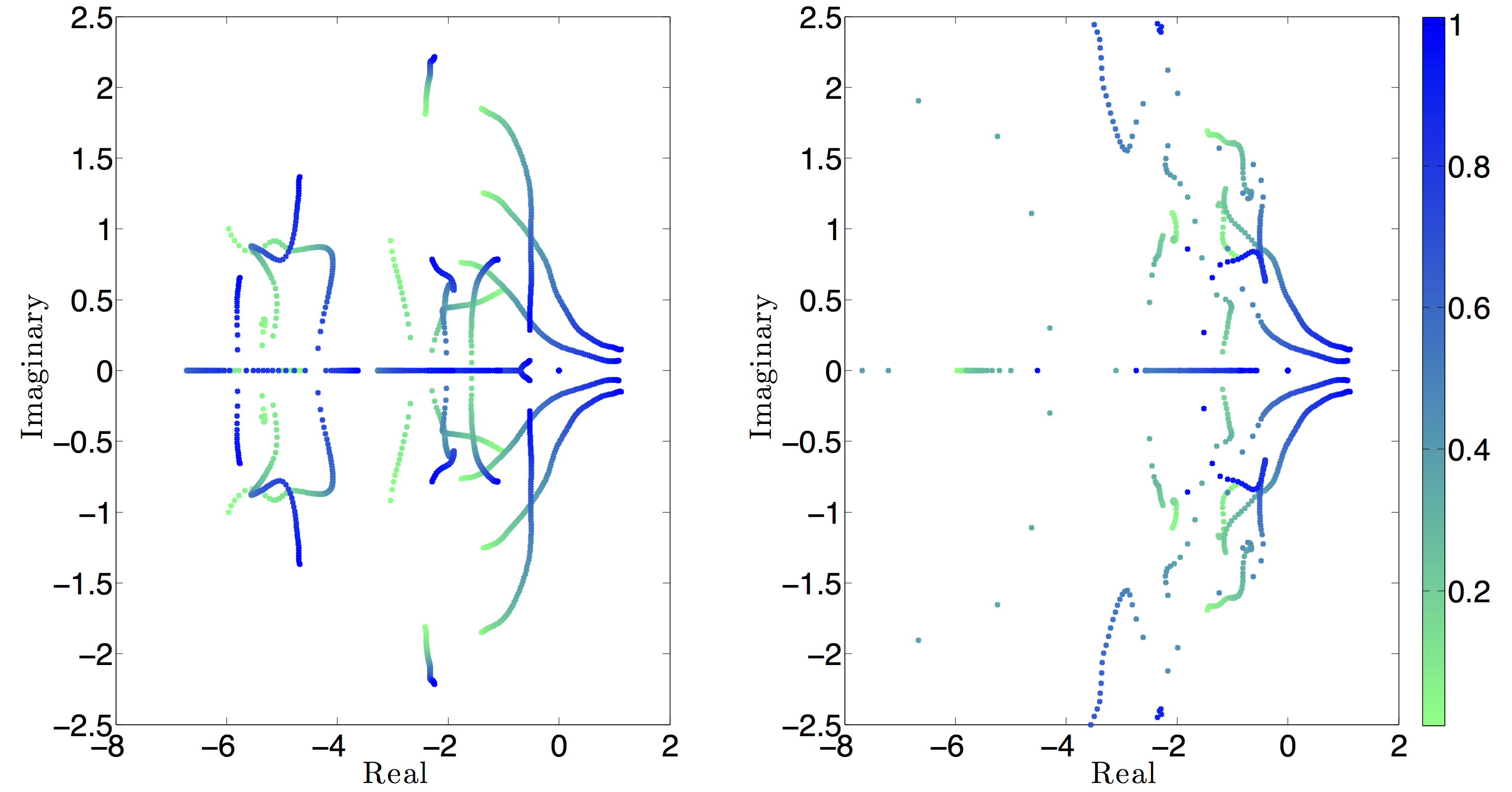}
\caption{\hl Pole migration map for the reduced order model (left) and for the collection of locally reduced systems (right). The value of the scheduling parameter is indicated by colors according to the color-bar on the right.} \label{Fig:polemaps}
\end{figure}

Finally, the proposed algorithm {\hl is} compared with a local reduction approach. At each grid point a balanced transformation based model reduction {\hl is} performed for the corresponding high dimensional LTI model. The local Hankel singular values imply lower dimensional models, within the range of $2-10$, for the sake of consistency the dimension has been fixed for $10$ at each grid-point. This fact indicates some conservativeness of the proposed methodology, which, in general preserves more states as a consequence of the modal decomposition. The pole migration maps for the two approaches are compared in Figure \ref{Fig:polemaps}. It can be observed, that the resulting parameter varying model has a smooth pole map, which makes {\hl the LPV model generation} straightforward and less challenging. On the other hand, the pole map for the set of locally reduced models shows large variation (see the right plot in Figure \ref{Fig:polemaps}). This implies the need of a more refined interpolation technique to successfully recover time domain behaviour of the original plant, which also makes control design virtually infeasible.
\end{itemize}

\subsection{B-1 aircraft}

Our second example is the publicly available nonlinear aeroelastic model of the Rockwell B-1 aircraft. This model is widely known and investigated due to the aeroelastic effects {\hl observed  at subsonic speeds}. {\hl To obtain an LPV system for model reduction the  simulation model from \cite{b1simulator} is used. By linearizing the nonlinear model at the flight altitude of $15000ft$ with different Mach numbers between $0.6-0.75$ grid-based LPV dynamics with $20$ state variables are obtained. The $20$ states consist of the $10$ dimensional rigid body dynamics (bank angle $\phi$, pitch angle $\theta$, yaw angle $\psi$, roll, pitch and yaw rates of the center of gravity: $p$, $q$, $r$, and $x-y-z$ axis velocities in the body coordinate frame: $U$, $V$, $W$ and altitude $h$) and the additional $10$ states of the five flexible modes.}  Eight measured outputs are considered: flight path angle  $\gamma$,  the roll, pitch and yaw rates of the center of gravity, and the lateral and vertical accelerations  of the cockpit and center of gravity. The model has $9$ inputs: throttle $T$, right and left (symmetric) horizontal tails: $\delta_{HR}$, $\delta_{HL}$, upper and lower split rudder surfaces: $\delta_{RU}$, $\delta_{RL}$, wing upper-surface spoilers: $\delta_{SR}$, $\delta_{SL}$ and canard control vanes: $\delta_{CVR}$, $\delta_{CVL}$. For these actuators a $13$ dimensional LTI dynamics is given and used through the augmentation of the LPV model. Accordingly, the final model used for testing the model reduction algorithm is $33$ dimensional.  

The reduced order model can be obtained by performing the following steps. After obtaining the parameter-varying modal form, three mixed stability modes {\hl are} removed. The resulting $30$ states {\hl can be} grouped into $3$ clusters with dimensions of $23$, $5$ and $2$. Based on the parameter-varying Gramians, the ASVD plots indicated $7$, $5$ and $2$  significant dimensions (i.e. only the largest block had been reduced). Hence, the resulting reduced-order model has a $14$ dimensional stable and $3$ dimensional unstable part.

\begin{figure}
\centering
\includegraphics[width=0.8\columnwidth]{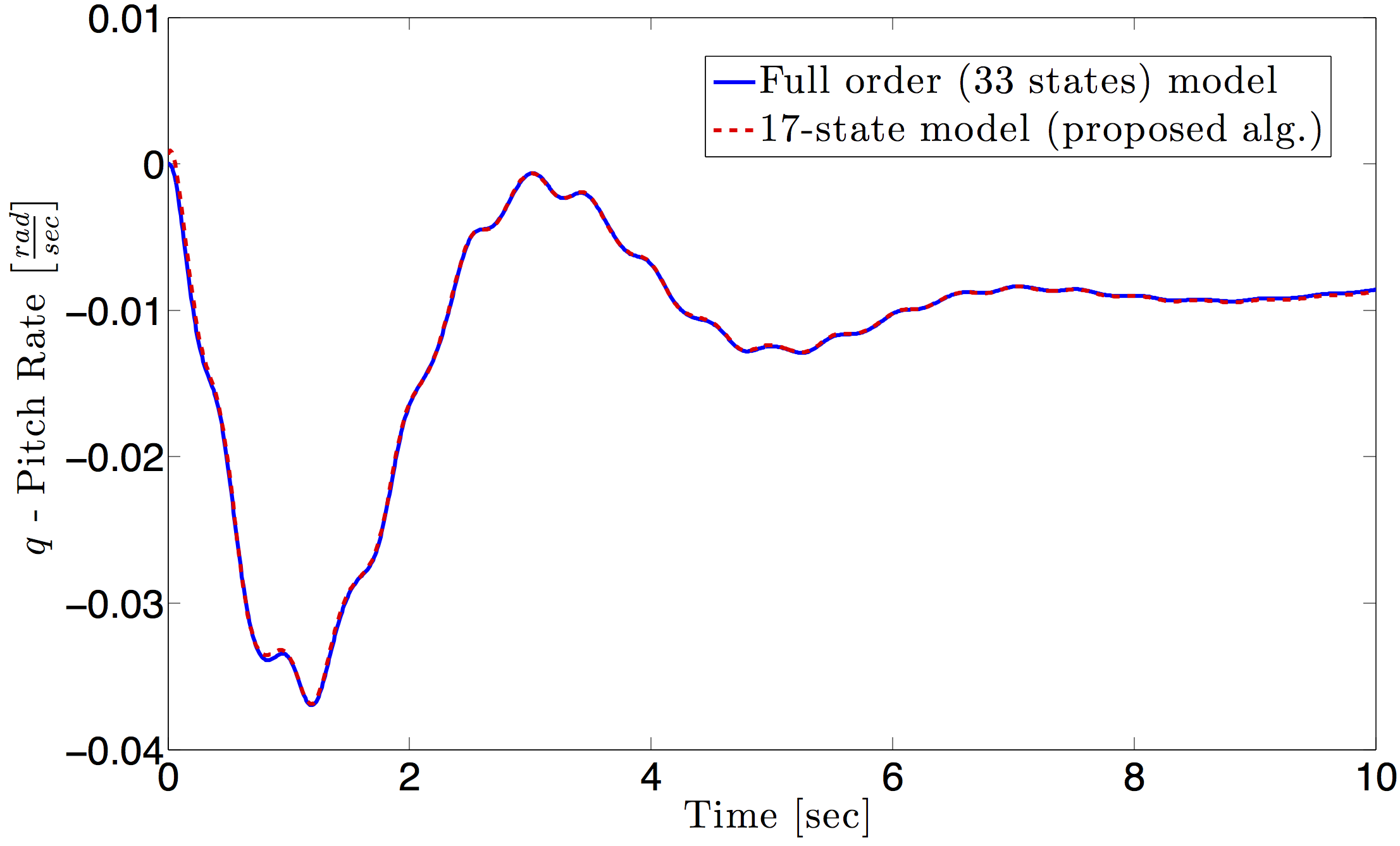}
\caption{Pitch rate response to a positive one degree deflection of the symmetric horizontal tail. \label{Fig:symmetricpitch}}
\end{figure}

\begin{figure}
\centering
\includegraphics[width=0.8\columnwidth]{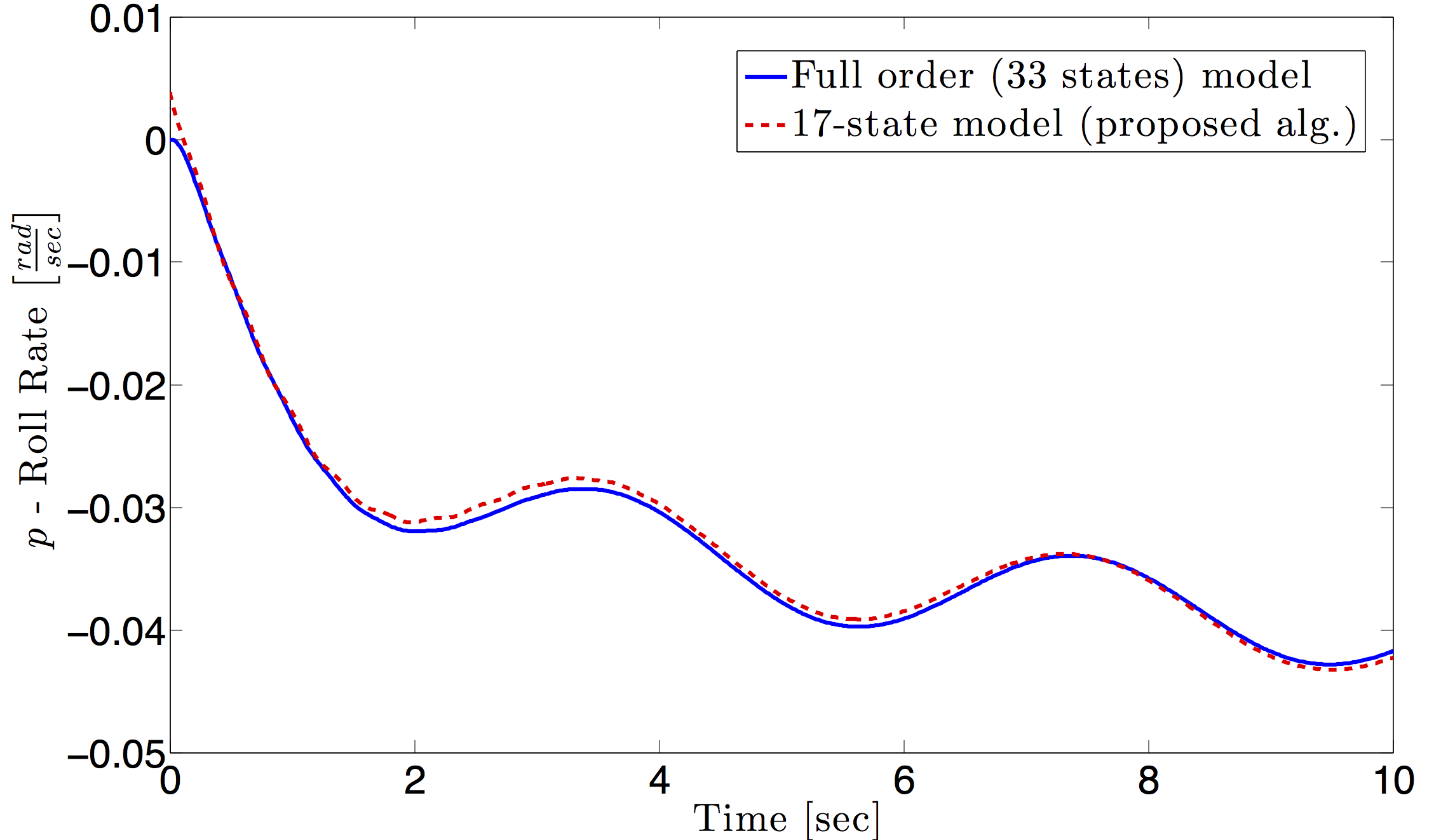}
\caption{Roll rate response to {\hl a one degree}  asymmetric horizontal tail deflection. \label{Fig:asymmetricroll}}
\end{figure}

{\hl To investigate the numerical properties of the reduced order model}, time-domain simulations are carried out first. In these simulations the flight speed {\hl is} set to vary as:
\begin{equation}
V(t)=0.675+0.05\sin(0.2t),
\end{equation}
which corresponds to a maximal acceleration of $\approx 3.3 \frac{m}{sec^2}$. Figure \ref{Fig:symmetricpitch} compares the pitch rate responses to a positive-one-degree commanded symmetric-horizontal-tail deflection of the full order model (with $33$ states) and the reduced order one ($17$ states). It can be depicted that the two models give almost identical responses. Figure \ref{Fig:asymmetricroll} shows the roll rate responses, where the horizontal-tail deflections are set asymmetrically $\delta_{HL}=-1^ {\circ}$, $\delta_{HR}=1^{\circ}$. Again, only a small discrepency can be observed between the two model. 

\begin{figure}
\centering
\includegraphics[width=0.9\columnwidth]{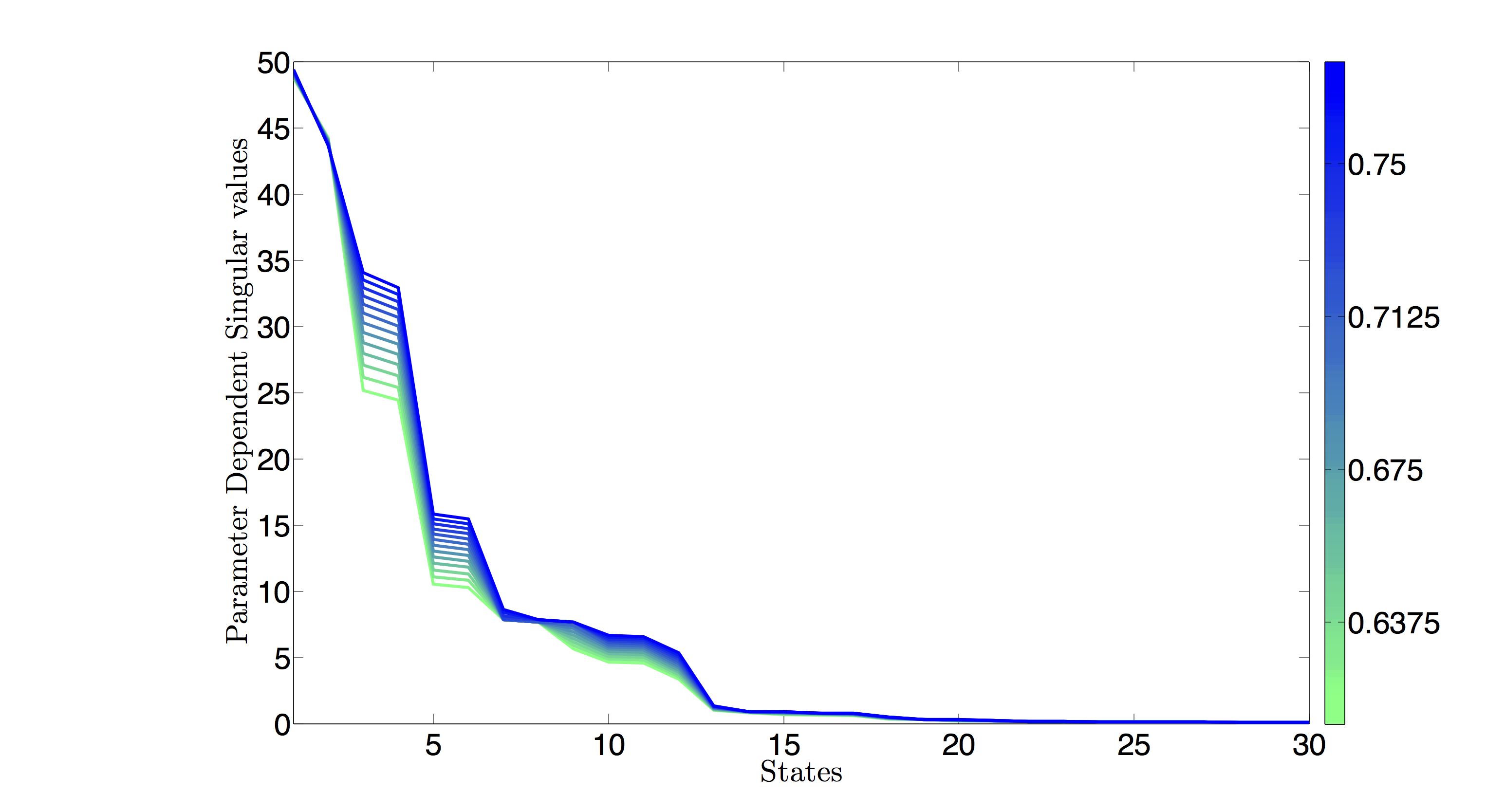}
\caption{Singular Values for the stable part of the B1 aircraft model.  The value of the scheduling parameter is indicated by colors according to the color-bar on the right. \label{Fig:hankel}}
\end{figure}

{\hl As noted previously, the LPV balanced transformation and reduction suffers from the curse of dimensionality, i.e. it runs into numerical problems for higher dimensional systems. Accordingly, it cannot be applied directly to } the benchmark example in Section \ref{subsec:benchmark}. At the same time, for the case of the $33$ dimensional model it is possible to solve the underlying LMI problems and consequently obtain a reduced description. Accordingly, after a stable-unstable decomposition the parameter-varying singular values have been computed for the stable part by using the parameter-varying Gramians and the corresponding balancing transformation \cite{wood95_phd}. The singular-value plots are given in Figure \ref{Fig:hankel} implying the elimination of $17$ states, which corresponds to a $16$ dimensional reduced system. This shows and verifies that the proposed approach indeed {\hl give} a good estimation for the order of the reduced system. 

\begin{figure}
\centering
\includegraphics[width=0.9\columnwidth]{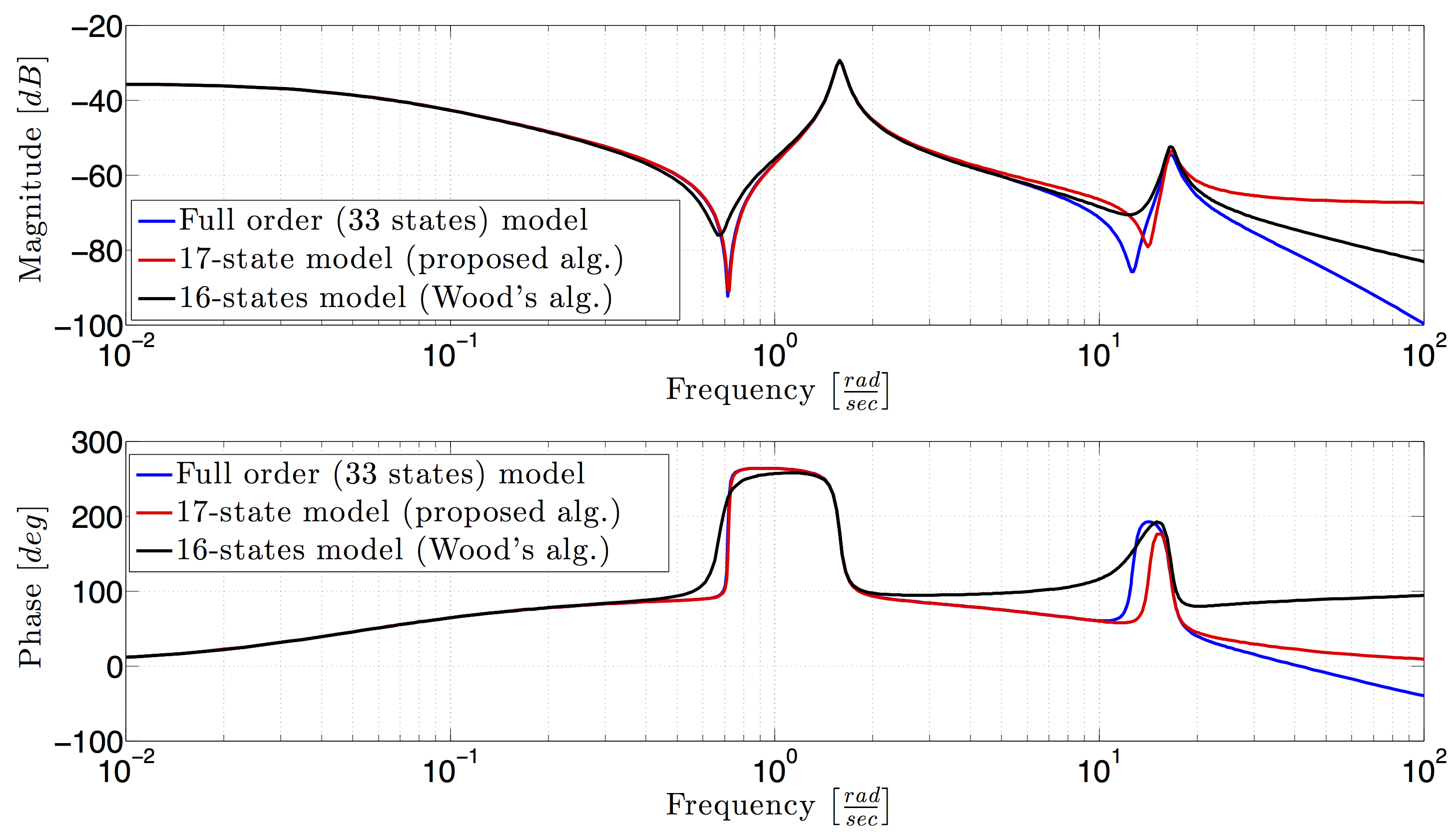}
\caption{Bode plots from $\delta_{RU}$ to $r$ of the full model ($33$ states) at $0.6$ Mach compared to  the two reduced order models. The $17$-state 
model has been obtained by the proposed algorithm, while the balanced reduction algorithm from \cite{wood95_phd} has generated the $16$-state model.  \label{Fig:R_u2r-M06}}
\end{figure}

\begin{figure}
\centering
\includegraphics[width=0.9\columnwidth]{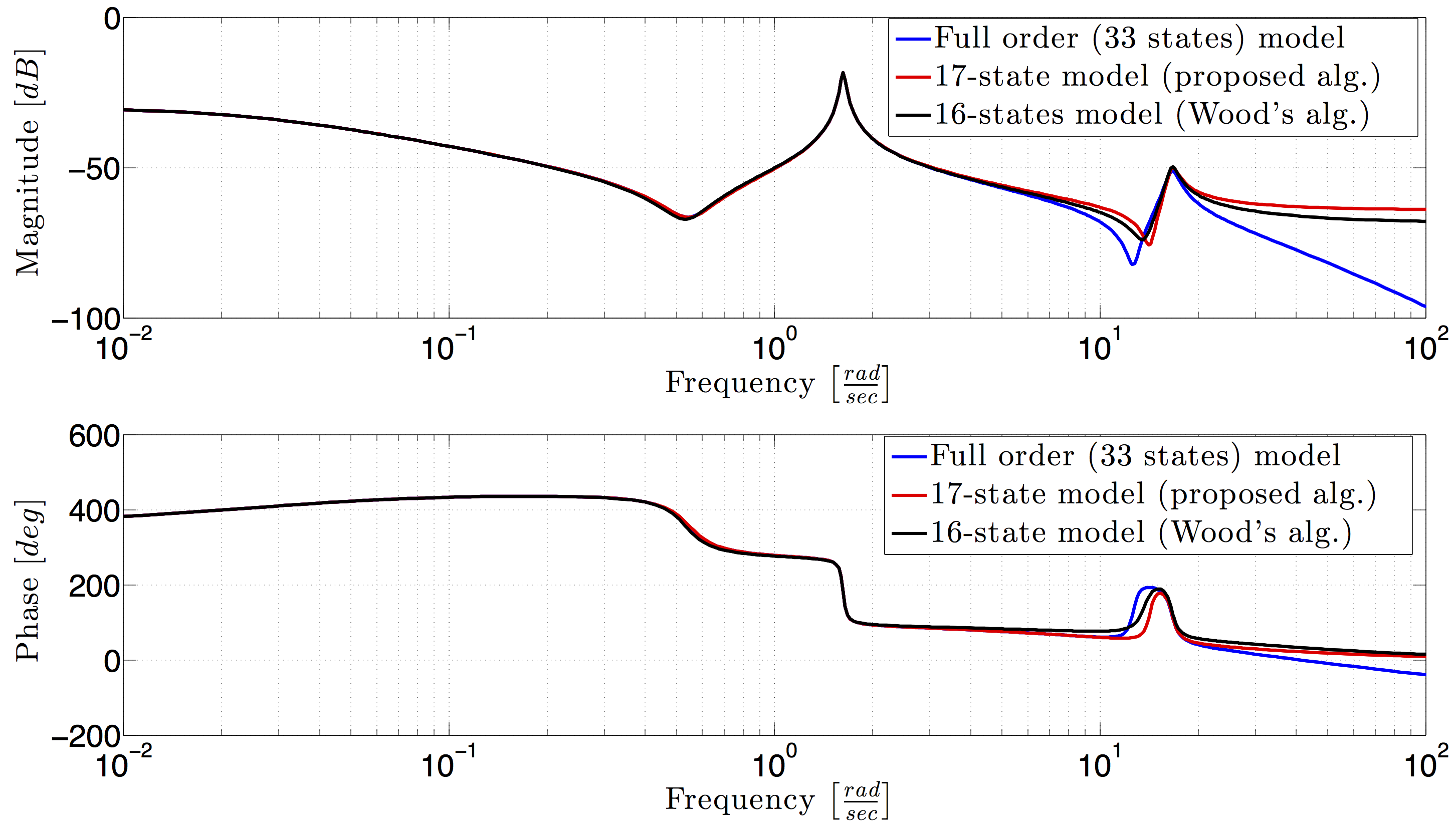}
\caption{Bode plots from $\delta_{RU}$ to $r$ of the full model ($33$ states) at $0.7312$ Mach compared to the two reduced order models. The $17$-state 
model has been obtained by the proposed algorithm, while the balanced reduction algorithm from \cite{wood95_phd} has generated the $16$-state model. \label{Fig:R_u2r-M075}}
\end{figure}

Finally, the frequency-domain properties are compared in Figures \ref{Fig:R_u2r-M06} and \ref{Fig:R_u2r-M075}. The Bode amplitude and phase diagrams of the i) full (blue), ii) reduced with the proposed approach (red) and iii) reduced by the method of \cite{wood95_phd} (black), models have been compared from the upper rudder surface $\delta_{RU}$ to the roll rate $r$ at two different flight speed ($0.6M$ and the interpolated point $0.7312M$). Note that, according to the applied balanced residualization, model mismatch appears in the higher frequency domain. Still, the frequency domain approximation is satisfactory in the operation domain. 

\section{Conclusion}
A novel model order reduction algorithm, based on approximate system decomposition, has been developed for LPV systems. The fundamental element of the algorithm is a parameter varying state transformation, which transforms (at least approximately) the LPV system into modal form. For this purpose the local transformation matrices have been appropriately modified and interpolated. The obtained structure is then subjected to a dynamic similarity analysis by using a hierarchical clustering method. This step reveals dynamically redundant modes of the system, which are grouped together to form parameter-varying subsystems.  Finally, these smaller dimensional subsystems are separately reduced by parameter varying balanced reduction algorithm. The effectiveness of the methodology is illustrated by in-depth numerical studies. 

{\hl The present framework requires some assumptions on the LPV system to be reduced. It is shown that most of these assumptions can be relaxed if slight modifications are allowed on the system matrices. The most stringent constraint is the one limiting the dimension of the scheduling variable to 1. Though the main concept of the proposed model reduction method can be extended to systems having more than one scheduling parameters, the numerical details of the algorithm have to be elaborated. The most critical point here is the separation of the modal subsystems, i.e. the generalization of the Hungarian algorithm and the iterative Procrustes smoothing. But once this is done the procedure can be continued since the clustering and the balanced reduction do not depend on the dimension of the scheduling parameter. The extension of the modal decomposition to multi-parameter LPV systems is one important direction for the future research. }

The well-known numerical problems related to eigen-decomposition is certainly one area which also needs further considerations, especially for very large dimensional and ill-conditioned problems. Furthermore, currently we are seeking more rigoruous methods for the decomposition of LPV systems. Guaranteed error bounds of the method will also be part of our future research.

\subsection*{Acknowledgement}
The research leading to these results is part of the FLEXOP project. This project has received funding from the European Unions Horizon 2020 research and innovation programme under grant agreement No 636307. 
The authors would like to thank Peter Seiler (University of Minnesota) and Arnar Hjartarson (formerly MUSYN Inc.) for providing insight related to the B-1 simulator model, originally developed by David Schmidt. {\hll The authors also would like to thank prof. Gary Balas (University of Minnesota) for initiating the research on control oriented modeling of flexible aircrafts, which gave motivation for working on large-scale LPV systems.}

\bibliographystyle{plain}
\bibliography{ijrnc_modred}
\end{document}